%% file: main.tex
\documentclass[letterpaper,twocolumn,10pt]{article}
\usepackage{hyperref}
\usepackage{usenix,epsfig,endnotes}

\newcommand{\PAGENUMBERS}{yes}       

\newcommand{\showComments}{yes}
\newcommand{\comment}[1]{}
\newcommand{\onlyAbstract}{no}
\newcommand{\makePasteable}{no}  

\usepackage{ifthen}
\usepackage{comment}

\usepackage[font=bf]{caption}
\usepackage{subcaption}

\usepackage{titlesec}

\usepackage{enumitem}

\usepackage{fancyhdr}

\ifthenelse{\equal{\PAGENUMBERS}{yes}}{%
  \pagestyle{plain}
}{%
  \pagestyle{empty}
}

\usepackage[numbers]{natbib}

\usepackage{booktabs}
\usepackage{color}
\usepackage{colortbl}
\usepackage{float}                           

\usepackage{mathptmx}                        
\usepackage{courier}
\usepackage[scaled=0.92]{helvet}

\definecolor{placeholderbg}{rgb}{0.85,0.85,0.85}

\ifthenelse{\equal{\makePasteable}{yes}}{%
\hyphenpenalty 10000
  \exhyphenpenalty 10000
  \setlength{\paperheight}{10740pt}
  \setlength{\textheight}{10740pt}
  \onecolumn
}{%
}

\usepackage{soul}
\usepackage{fancyvrb}
\usepackage{multirow}
\usepackage{rotating}
\usepackage{tikz}
\usepackage{stfloats}
\usepackage{cuted}
\newcommand*{\circled}[1]{\lower.7ex\hbox{\tikz\draw (0pt, 0pt)%
    circle (.5em) node {\makebox[1em][c]{\small #1}};}}

\usepackage{hhline}
\usepackage{pgfplots}
\usepackage{pgfkeys}
    \def\addlegendimage{\csname pgfplots@addlegendimage\endcsname}
\pgfplotsset{
cycle list={%
{draw=black,mark=star,solid},
{draw=black, mark=square,solid},
{draw=black,mark=+,solid},
{black,mark=o},}}
\usepackage[absolute]{textpos}
\usepackage[yyyymmdd]{datetime}

\newcommand{\Timestamp}{%
\begin{textblock*}{\textwidth}(0.65in,0.4in)
\begin{flushleft}
     \noindent{\textsf{\date{\today\ @ \currenttime}}}
\end{flushleft}
\end{textblock*}
}

\newcommand{\ToAppear}{%
\begin{textblock*}{\textwidth}(0.95in,0.4in)
\begin{flushright}
\end{flushright}
\end{textblock*}
}

\newfloat{acmcr}{b}{acmcr}

\usepackage{authblk}
\usepackage[export]{adjustbox}
\usepackage{xspace}
\usepackage{algorithm2e}
\usepackage{makecell, multirow}
\setcellgapes{3pt}
\newcommand{\note}[2]{
    \ifthenelse{\equal{\showComments}{yes}}{{\color{#1}[#2]}}{}
}

\newcommand{\commentcolor}[2]{
  \ifthenelse{\equal{\showComments}{yes}}{{\color{#1}#2}}{#2}
}

\definecolor{eblue}{RGB}{0, 0, 139}
\definecolor{mblue}{RGB}{0, 147, 175}
\definecolor{rgreen}{RGB}{0, 112, 60}
\definecolor{worange}{RGB}{245, 128, 37}
\definecolor{dblue}{RGB}{0, 0, 255}

\newcommand{\textblue}[1]{\textcolor{eblue}{#1}}
\ifx\noeditingmarks\undefined
   \newcommand{\pgwrapper}[2]{\textblue{#1: #2}}
\else
   \newcommand{\pgwrapper}[2]{}
\fi

\newcommand{\name}{MAT\xspace}

\date{}
\title{\textbf{Machine Learning Over Heuristic:\\ a Learned Cache Eviction Framework with Minimal Overhead}}
\author[1]{Dongsheng Yang}
\author[2]{Daniel S. Berger}
\author[1]{Kai Li}
\author[2]{Wyatt Lloyd}
\affil[1]{Princeton University}
\affil[2]{Microsoft Research}

\begin{document}

\maketitle

\Timestamp
\ToAppear

\input{abstract}
\ifthenelse{\equal{\onlyAbstract}{no}}{%
\input{intro}
\input{background}
\input{design}

\input{eval}
\input{related}

\input{concl}

\setlength{\bibsep}{2pt}
\small 
\bibliography{ref}
\bibliographystyle{abbrvnat}

\appendix

}{
}

\end{document}

%% file: abstract.tex
\begin{abstract}
Recent work shows the effectiveness of Machine Learning (ML) to reduce cache miss ratios by making better eviction decisions than heuristics.  However, state-of-the-art ML caches require many predictions to make an eviction decision, making them impractical for high-throughput caching systems.


This paper introduces Machine learning At the Tail (\name{}), a framework to build efficient ML-based caching systems by integrating an ML module with a traditional cache system based on a heuristic algorithm.
\name{} treats the heuristic algorithm as a ``filter'' to   
receive high-quality samples to train an ML model and likely candidate objects for evictions. 
We evaluate \name{} on 8 production workloads, spanning storage, in-memory caching, and CDNs. The simulation experiments show \name{} reduces the number of costly ML predictions-per-eviction from 63 to 2, while
achieving comparable  miss ratios to the 
state-of-the-art ML cache system.  We compare a \name{} prototype system with an LRU-based caching system in the same setting and show that achieve similar request rates.


\end{abstract}

%% file: intro.tex
\section{Introduction}
\label{sec:intro}

Software caching systems are ubiquitous
in modern computing infrastructure.  Examples of large-scale use cases include
include content delivery networks (CDNs), in-memory caches, and storage systems.
CDNs protect expensive and scarce Internet backbone bandwidth and are expected to serve 72\% of Internet traffic by 2022~\cite{vni2019}.
In-memory caches protect computationally expensive services are extensively used in the data centers of Facebook~\cite{mcallister2021kangaroo} and Twitter~\cite{yang2020large}.
Storage caches reduce the data movement of large objects in the network and an essential part of cloud services~\cite{eytan2020s}.

Caching systems seek to minimize their miss ratio, i.e., the fraction of requests not served by the cache.
The lower the miss ratios, the lower the load on backend servers and Internet traffic (for CDNs).
To decide which objects to keep in the cache, current caching systems~\cite{redis, memcached, berg2020cachelib,mcallister2021kangaroo} rely on heuristic algorithms, such as Least Recently Used (LRU), and First In First out (FIFO), and Least Frequently Used (LFU).
Recent work~\cite{song2020learning,wang2019learning,akhtar2019avic,yan2020rl,berger2018towards} shows that machine learning based eviction algorithms (ML-based caching systems) significantly outperform these heuristics by using a history of past access patterns to predict future access patterns.
These accurate predictions reduce miss ratios by up to 25\% compared to heuristic caches~\cite{song2020learning}.

Bringing ML-based caching systems from research to production faces a key challenge due to their computational overhead and hardware cost.
In particular, ML-based caching systems are not yet applicable in systems with high throughput demands~\cite{beckmann:nsdi18:lhd,fan2013memc3} or when CPU resources are scarce due to being coloated with other applications~\cite{berg2020cachelib}.

The overhead of ML-based caches is significantly higher than heuristic caching systems for two reasons.
First, ML-based caching systems need to update the model online frequently to retrain with more recent access patterns. For example, a state-of-the-art ML-based caching system for CDNs, LRB~\cite{song2020learning}, uses all cache requests to generate training entries, which leads to a large training data volume and a slow training process. 

Second, ML-based caches require running many predictions to find an object to evict.
For example, LRB samples $64$ eviction candidates randomly within the cache to run predictions.
Running 64 predictions per eviction can be slow and expensive especially in bursty production systems that can face pressure to evict hundreds of thousands of evictions in a second due to burst arrivals.

While the overheads of ML-based caches are known, it is less known which of their decisions are actually required to improve miss ratios compared to heuristics.
An answer to this question can guide applying costly ML predictions only where they are needed.
In fact, when comparing the eviction decisions of heuristics to an offline optimal algorithm, we find that they evict most of the objects that the optimal algorithm evicts, but they sometimes evict objects they should keep.
This leads to our main insight that heuristic algorithms can serve as good filters. The ML algorithm will only run predictions on the objects evicted by the heuristic algorithm, instead of all objects in the cache. It can dramatically reduce the number of predictions without affecting miss ratios.  

We propose an efficient ML-based caching framework, \textit{Machine learning At the Tail} (\name{}), which builds on the insight that we can effectively pair a heuristic with an ML predictor.
We define the \emph{tail} as evictions of the heuristic algorithm, e.g., the least recently used items in LRU.
\name{} feeds the tail objects into a novel ML predictor.
This ML predictor then decides which objects to keep in the cache and which objects to truly evict.
This allows \name{} to identify good objects to evict using only a few predictions because the heuristic's tail is a small subset of objects in the cache. Similarly, it allows \name{} to focus its training on this small subset of objects.
In turn, this means \name{} does far less computation per eviction than prior ML-based caching system.
But, because the heuristic's tail contains nearly all the objects that should be evicted, \name{} can achieve the same miss ratio as state-of-the-art ML-based caching systems.




An additional challenge in many caches is handling scenarios where computation power becomes scarce during certain time intervals such as request load spiking or an increase in higher-priority work that is collocated with the cache~\cite{berg2020cachelib}.
MAT's design robustly handles these scarce computation scenarios by falling back to the heuristic algorithm when the ML model cannot keep up with the request load.

To compare \name{}  with a variety of algorithms including LRB, the state-of-the-art ML cache,
We have also implemented it in a cache simulator and run experiments with 8 production workloads from CDNs, in-memory caches, and storage caches.  Our results show that while achieving comparable miss ratios, \name{} dramatically reduces the overhead for ML:
it reduces the average number of predictions per eviction by 31 times (from 63 to 2) and the average prediction overhead per eviction including metadata feature building overhead by 21 times (from 300us to 9.3us).

We have implemented \name{} in Cachelib~\cite{berg2020cachelib}, which is an open-source cache system developed by Facebook.
We compare its performance with the Cachelib instance with LRU algorithm.  
Our end-to-end evaluation shows that \name{} achieves similar request rates to the LRU-based caching system. 


Section~\ref{sec:motivation} elaborates on the motivation of our work.  Section~\ref{sec:insight} details the observation that leads to our design. In Section~\ref{sec:design} we present the design of our framework \name{}. Section~\ref{sec:implementation} describes how MAT is implemented in the simulator and in the Cachelib prototype.
Section~\ref{sec:eval} presents an evaluation of \name{}. We
cover related work in Section~\ref{sec:related} and we conclude in
Section~\ref{sec:concl}.

%% file: background.tex
\section{Background and Motivation}
\label{sec:motivation}

This section covers background on offline caching algorithm, heuristic caching algorithms, and ML caching algorithms.
We will use examples from each group to study the decision quality of heuristic algorithms in Section~\ref{sec:insight}.

\subsection{Optimal Offline Algorithm}

In 1966, Belady proposed the MIN algorithm that evicts a data object whose next access occurs furthest in the future~\cite{belady}.  This algorithm provably optimal for caching equal-sized data objects~\cite{mattson} such as cacheline or video chunks.
Since knowledge about future accesses is not typically available in an online setting, we call Belady's MIN algorithm an \emph{offline optimal} or oracle algorithm.

\subsection{Heuristic Caching Algorithms}

\begin{figure}[!htbp]
    \centering
    \includegraphics[width=0.95\linewidth,valign=t]{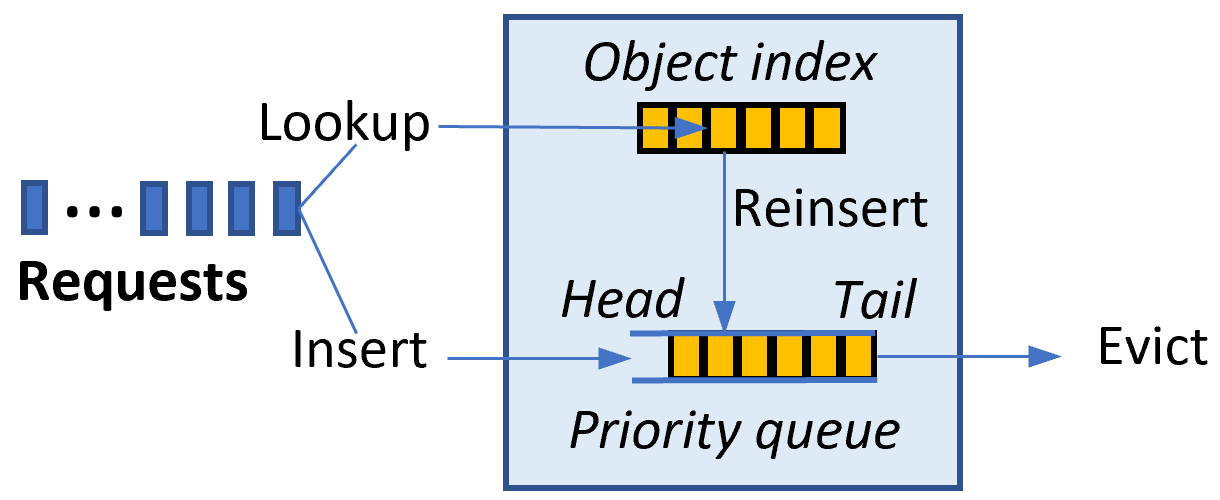}
    \caption{Heuristic cache algorithms that maintain the rank of objects in a priority queue.}
    \label{fig:heu}
\end{figure}

The most common class of caching algorithms used in production systems are based on 
heuristics.  A heuristic is designed to decide which object to evict from a cache to admit an object that is currently not cached.
Most heuristic algorithms form an explicit or implicit ranking of objects in the cache. If the cache request is a miss, it inserts the requested object into the ranking and evicts the object with the lowest rank. If the cache request is a hit, it re-ranks the requested object in the cache.
Figure~\ref{fig:heu} shows the typical structure of a heuristic caching algorithm.

A well-known example is Least-Recently-Used (LRU), which maintains a queue to implicitly rank objects by their most recent access times. LRU inserts the most recently accessed object at the head of the queue 
and evicts the one at the tail of the queue.


The main advantage of heuristic caching algorithms is their simplicity and efficiency.  For example, LRU can be implemented with a doubly-linked list as its priority queue, and a hash table to speedup the lookup operation.
This implementation of LRU is the default algorithm in many production systems such as Cachelib~\cite{berg2020cachelib}.

The main drawback of heuristic caching algorithms is that they  work well for certain workloads or access patterns while working poorly for others~\cite{song2020learning}.





\subsection{ML-Based Caching Algorithms}
\label{sec:lrb}

Machine learning is changing how caches are designed.  An ML-based cache trains a model with past access patterns and then uses the model to predict which objects in the cache should be evicted.
Recent studies~\cite{berger2018towards, song2020learning, zhou2021learning, yan2021learning} 
show that ML-based approaches can adapt to different workloads dynamically and can reduce wide area network traffic by around 20\% compared to the state-of-the-art heuristic algorithms.  

\paragraph{Two Key Challenges.}
There are two key challenges with ML-based caching systems.  The first is the overhead for training ML models.  Adapting to recent access patterns requires training and updating the model frequently. This overhead can be significant in space and time as hardware accelerators are usually not equipped on caching servers.

The second is the overhead for making eviction decisions.
To mimic the optimal offline oracle in a straightforward way, the ML-based caching system needs to predict the next access times of all objects in the cache and evicts the one with the furthest time in the future. The prediction overhead would be prohibitively high for large caches. Therefore, ML-based caching systems for software caching have to decide which subset of objects to run predictions on.


\paragraph{ML-Based Caching by Sampling.}

\begin{figure}[!htbp]
    \centering
    \includegraphics[width=0.85\linewidth,valign=t]{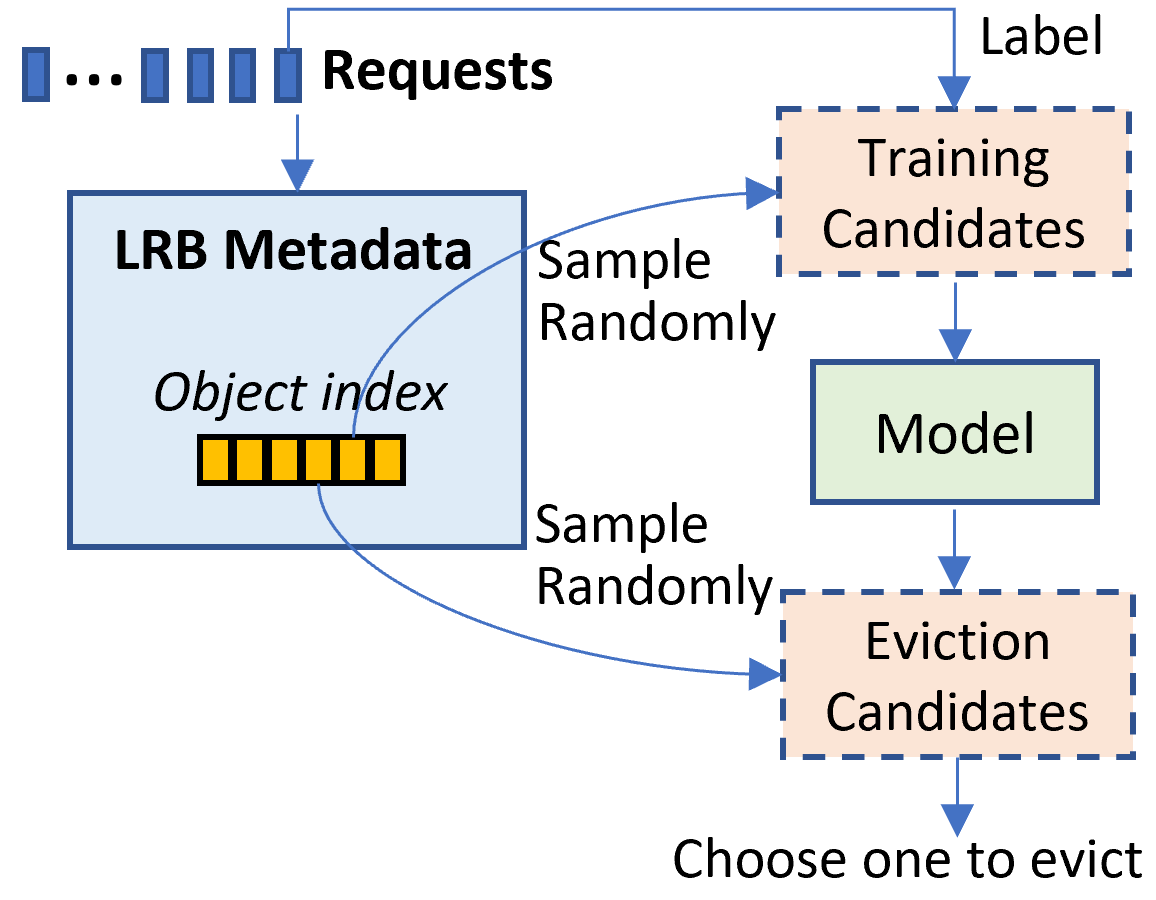}
    \caption{The state-of-the-art ML-based caching system, Learning Relaxed Belady (LRB).}
    \label{fig:lrb}
\end{figure}

LRB~\cite{song2020learning} is the state-of-the-art ML-based caching system and it is based on sampling for both training and for eviction selection.  
Figure \ref{fig:lrb} shows how it trains the ML model and uses the ML model to make evictions.

LRB overcomes the first challenge by using a Gradient Boosted Decision Tree (GBDT), a relatively simple ML approach. 
It trains and updates the GBDT model online with a relatively small training dataset (about 128K objects) randomly sampled from a window of recent request history.
The training overhead is small enough to update the model every few seconds.

It overcomes the second challenge by relaxing the eviction criteria of the Belady's MIN algorithm.  Instead of finding the object in the cache whose next access time occurs the furthest in the future, it picks any object whose predicted next access time is far enough.
With this approximation, LRB approach runs predictions on only $k$ randomly sampled objects in the cache, where $k$ is set to 64. When $k=64$, with a high probability, LRB can find at least one object whose predicted next access time is close to the largest next access time in the cache.


LRB achieved better miss ratios than 14 state-of-the-art heuristic caching algorithms over various cache sizes with 6 production CDN workloads.  Since LRB is designed for CDN workloads whose objects are quite large and requests rates are low, it can afford some computational overhead.
However, we find that LRB requires more than two orders of magnitude more CPU resources than heuristic algorithms.
We seek to reduce this overhead to enable us to deploy ML-based algorithms in high-throughput environments, application-embedded environments, or low-power environments such as the Internet edge.



\paragraph{Insertion-time ML caching algorithms.}
Another category of ML caching algorithms runs a prediction on each object as its request arrives~\cite{berger2018towards}. It maintains a data structure that remembers the ranking of the predicted scores of all objects in the cache and chooses the lowest ranked object for an eviction. 

The prediction overhead of this method is significant for two main reasons.  
First, the number of requests can be larger than the number of evictions by an order of magnitude, depending on cache miss ratios.  Second, after updating the ML model with new training data, the previous predictions are not consistent with the new model.  
It needs to rerun predictions of all objects in the cache to make the ranking up-to-date and consistent \cite{zhou2021learning}.  If the frequency of retraining the model is high, the total cost for predictions can be extremely high.

ML-based caching systems make better eviction decisions and thus they can save cost on storage media and network bandwidth. However, ML-based caching systems require a lot more computation than heuristic algorithms. In high throughput caching systems, the available computation power is not enough for the ML-based caching systems to keep up with the line speed. In other cases, it is also highly preferred to reduce the computational overhead of ML-based caching systems to save up power budget for other applications such as edge computing. Thus, our goal is to have as good eviction decisions as the state-of-the-art ML-based caching systems, while have orders of magnitude lower computational overhead.

\section{Heuristic Algorithms as Filters}
\label{sec:insight}

The key idea in this paper is to use a heuristic caching algorithm as a filter in front of an ML-based caching system to reduce the predictions per eviction and the samples for training an ML model. The question is how good heuristic algorithms are as such filters.

To answer this question, we compare the distribution of evicted objects by a heuristic algorithm to Belady's MIN (optimal offline) algorithm.  A good filter should pass over most good eviction candidates that Belady's MIN evicts, even at the cost of passing over some bad candidates.  We will first look at LRU algorithm as a filter and then look at several other heuristic algorithms.

\begin{table*}[!tb]
\centering
\begin{tabular}{|c|cc|cc|cc|cc|cc|}
\hline
     & \multicolumn{2}{c|}{LRU}                                  & \multicolumn{2}{c|}{FIFO}                                 & \multicolumn{2}{c|}{LFUDA}                                & \multicolumn{2}{c|}{LRUK}                                 & \multicolumn{2}{c|}{Belady's MIN}                           \\ \hline
     & \multicolumn{1}{c|}{TTA\textless{}\textit{T}} & TTA>\textit{T} & \multicolumn{1}{c|}{TTA\textless{}\textit{T}} & TTA\textgreater{}\textit{T} & \multicolumn{1}{c|}{TTA\textless{}\textit{T}} & TTA\textgreater{}\textit{T} & \multicolumn{1}{c|}{TTA\textless{}T\textit{}} & TTA\textgreater{}\textit{T} & \multicolumn{1}{c|}{TTA\textless{}\textit{T}} & TTA\textgreater{}\textit{T} \\ \hline
CDN1 & \multicolumn{1}{c|}{55\%}            & 90\%               & \multicolumn{1}{c|}{65\%}            & 90\%               & \multicolumn{1}{c|}{47\%}            & 87\%               & \multicolumn{1}{c|}{363\%}           & 97\%               & \multicolumn{1}{c|}{0\%}             & 100\%              \\ \hline
CDN2 & \multicolumn{1}{c|}{47\%}            & 95\%               & \multicolumn{1}{c|}{57\%}            & 95\%               & \multicolumn{1}{c|}{36\%}            & 96\%               & \multicolumn{1}{c|}{487\%}           & 92\%               & \multicolumn{1}{c|}{0\%}             & 100\%              \\ \hline
CDN3 & \multicolumn{1}{c|}{42\%}            & 94\%               & \multicolumn{1}{c|}{67\%}            & 91\%               & \multicolumn{1}{c|}{23\%}            & 95\%               & \multicolumn{1}{c|}{41\%}            & 96\%               & \multicolumn{1}{c|}{0\%}             & 100\%              \\ \hline
Wikipedia & \multicolumn{1}{c|}{129\%}           & 94\%               & \multicolumn{1}{c|}{196\%}           & 87\%               & \multicolumn{1}{c|}{86\%}            & 95\%               & \multicolumn{1}{c|}{89\%}            & 93\%               & \multicolumn{1}{c|}{0\%}             & 100\%              \\ \hline
\end{tabular}
\caption{Fractions of evicted objects whose TTA < $T$ and  TTA > $T$ by 5 caching algorithms (LRU, FIFO, LFUDA, LRUK and Belady's MIN) with 4 workloads (CDN1, CDN2, CDN3 and Wikipedia). All fractions are normalized to the total number of objects evicted by Belady's MIN. 
}
\label{tbl:eviction_analysis}
\end{table*}

\subsubsection*{Compare LRU to Belady's MIN}

\begin{figure}[!htbp]
    \centering
    \includegraphics[width=\linewidth,valign=t]{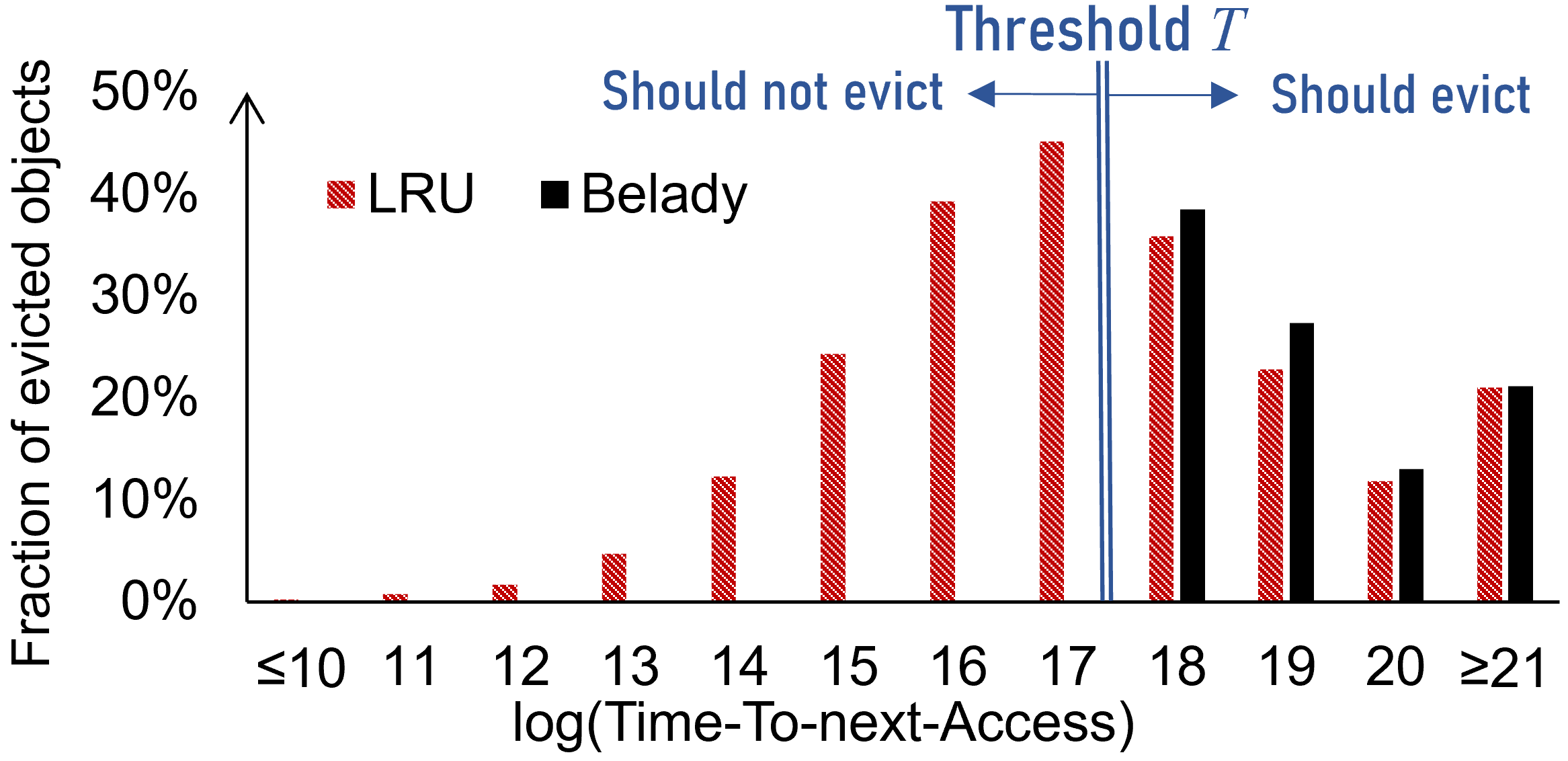}
    \caption{Time-To-next-Access (TTA) distribution of the evicted objects by LRU and Belady's MIN algorithms. The result is collected from the Wikipedia trace with 256GB cache size.
    }
    \label{fig:motivation}
\end{figure}

Figure~\ref{fig:motivation} compares the distributions of evicted objects by LRU and Belady's MIN for Wikipedia workload with a cache size of 256\,GB. 
The figure groups evicted objects according to the log of their Time-To-next-Accesses (TTAs) from the time when they are evicted by a given algorithm. The time is the logical time, which means it is a counter that increments on each request by 1. The figure plots the percentage of evictions in each group normalized by the total number of evictions of the Belady's MIN algorithm.

The threshold $T$ in the figure separates good eviction decisions from bad ones. 
All evicted objects by Belady's MIN have their TTAs $> T$ (on the right hand side of the threshold), none have TTAs $<T$ (on the left hand side of threshold). 

We have two observations about the distributions of LRU. 
First, the total number of evicted objects whose TTAs $>T$ by LRU is close to that by Belady's MIN. This means that LRU evicts most of objects that Belady's MIN does.  
In other words, LRU rarely keeps objects it should evict.
This indicates that in most cases, LRU does not filter out most of the good eviction candidates.  

Second, the number of objects evicted by LRU on the left hand side of the threshold is similar to that on the right hand side.  In other words, although LRU frequently evicts objects it should keep, one of every two evictions is a good decision on average for this workload.  This means that using LRU as a filter, we can reduce the number of predictions from 64 to 2!

\subsubsection*{Compare Other Heuristics to Belady's MIN.}


Table~\ref{tbl:eviction_analysis} shows the distributions of good and bad eviction decisions with four workloads by LRU, FIFO, LFUDA, LRUK and Belady's MIN algorithms. As Belady's MIN is an optimal offline algorithm,  100\% of its evicted objects have TTA $>T$. 

The main observation is that all heuristic algorithms in the table can serve as filters well.  They can evict 87-97\% of the objects (TTA $>T$) Belady's MIN evicts.  

Among these heuristic algorithms as filters,  LRU and LFUDA are better overall.  LRU evicts 90\%, 95\%, 94\%, and 94\% of the objects that Belady's MIN evicts with CD1, CDN2, DDN3 and Wiki workloads respectively.  LFUDA evicts 87\%, 96\%, 95\%, and 95\% of those that Belady's MIN evicts respectively.  LFUDA has the smallest fractions (23-86\%) of evicted objects whose TTA $<T$.

LRUK achieves the highest coverage (92-97\%) of the objects that Belady's MIN evicts, but it has large fractions of bad eviction decisions with CDN1 and CDN2 worklaods (363\% and 487\% respectively). 

FIFO can also be a good filter but it is slightly worse for Wiki workload, evicting 87\% of the objects that Belady's MIN evicts.


\subsubsection*{Filtering Training Samples}

Using a heuristic caching algorithm as a filter can deliver better training samples than random sampling.  Our key insight is that since the ML-based caching algorithm 
takes objects from the tail of the heuristic algorithm as eviction candidates, the ML model needs to learn only from such historical candidates to make better predictions.  



%% file: design.tex
\section{Machine Learning at the Tail }
\label{sec:design}

\begin{figure*}[t]
    \centering
    \includegraphics[width=0.9\textwidth]{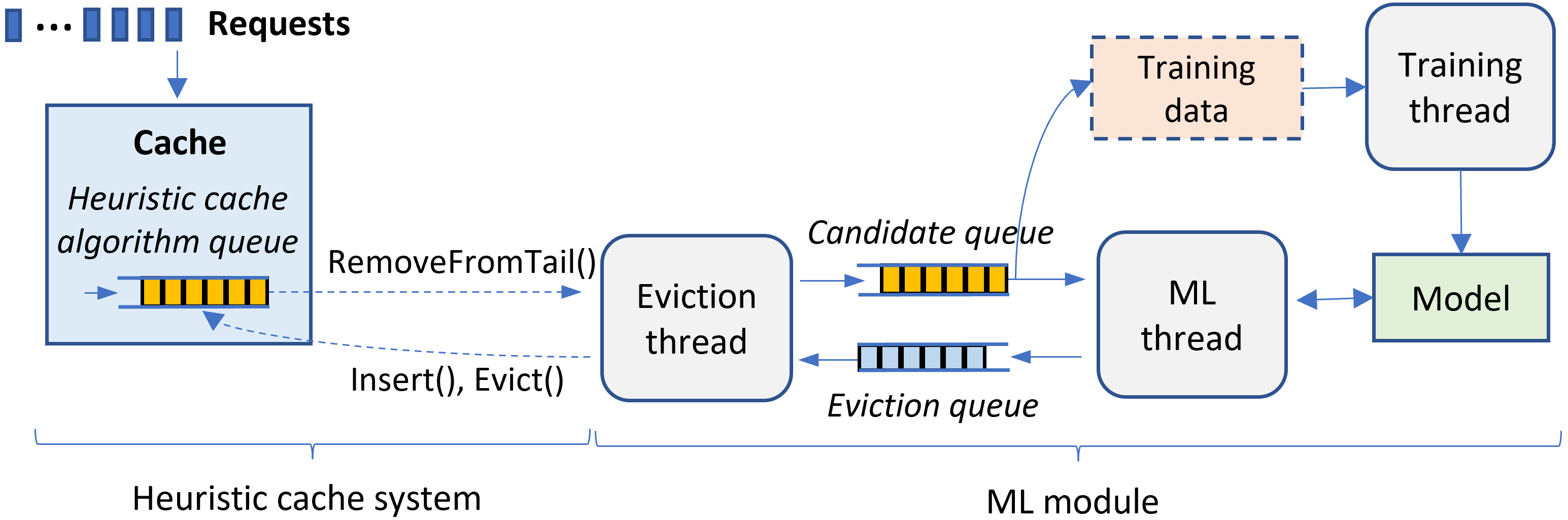}
    \caption{The architecture and the data flow of the \name{} system. \textmd{The ML module makes predictions on objects removed from the tail of the heuristic cache and decides to insert the objects back or evict the objects.}}
    \label{fig:designspace}
\end{figure*}

This section describes our approach called Machine Learning at the Tail (\name).  Section \ref{sec:arch} describes the two components of MAT, which are the heuristic cache system and the ML module, and their interfaces. Section \ref{sec:sample} describes and discusses how MAT uses a dynamic threshold to decide which object should be evicted. Section \ref{sec:mlmodel} introduces an implementation of the machine learning method of MAT. Finally, Section \ref{sec:mlmodel} describes how training data are generated in MAT.

\subsection{Architecture and Algorithm}
\label{sec:arch}

Figure \ref{fig:designspace} shows
the architecture of \name{}, consisting of two main modules: a heuristic cache  and an ML module.

\paragraph{Heuristic cache.}
It is a traditional cache using a priority-queue based heuristic algorithm.  It needs to provide two calls to interface with the ML module:
\begin{itemize}
    \item \texttt{RemoveFromTail}(): removes an object from the the tail of the priority queue(s) of the heuristic algorithm and return the object to the ML model.
    \item \texttt{Insert}($x$, $rank$): insert object $x$ back to a priority queue of the heuristic algorithm. The ML model can inform the heuristic algorithm to insert to a specific position of the queue by providing the rank.
\end{itemize}
In the case that the heuristic algorithm uses multiple queues, such as Segmented LRU, 2Q, and TinyLFU, the two procedures need to pick one of the queues.

The \name{} framework is general since most heuristic algorithms use priority queue(s).  We have implemented and experimented \name{} with LRU, 2Q~\cite{johnson19942q} and TinyLFU~\cite{einziger2014tinylfu} algorithms. The two calls are simple to implement.

\paragraph{ML Module.}
In the \name{} design, it consists of a training pipeline and a prediction (or inference) pipeline.
The prediction pipeline implements 
\texttt{Evict}() which returns an object for eviction.

The main data structure in the training pipeline is a training dataset, which is the recent historical candidates from the candidate queue.  The training thread uses the dataset to train a model and update the current model with the newly trained model. 

Two kinds of threads are used in the prediction pipeline: ML threads and eviction thread, connected by two  queues: candidate queue and eviction queue.    

An ML thread removes a batch of candidate objects from the candidate queue, predicts the time-to-next-access (TTA) of each object in a way similar to that of LRB~\cite{song2020learning}. 
If the TTA is greater than threshold $T$, the object will be put on the eviction queue. 

    
The eviction thread is responsible for removing objects from the tail of the heuristic algorithm
    (by calling \texttt{RemoveFromTail}())
    and putting them in the candidate queue, and insert them back into the priority queue(s) (by calling \texttt{Insert}()). 

When the cache system needs to evict an object from the cache, it will call \texttt{Evict}() which will remove an object from the eviction queue and return it as the object for eviction.  
If the eviction queue is empty, \texttt{Evict}() 
will remove an object at the tail of priority queue(s) of the heuristic algorithm.  
In either case, the cache system will evict the returned object from the cache and also removes it from its related priority queue. 

The main advantage to allow the cache system to go ahead when the eviction queue is empty is that the cache system can run at a speed (or throughput) similar to the heuristic cache system with a ML module.  In this case,  eviction decisions can fall back to the heuristic algorithm.  The cache system continues functioning well when the ML thread is slow or even fails.   

\subsection{Eviction Decision}
\label{sec:sample}

\RestyleAlgo{ruled}
\begin{algorithm}
\caption{MAT Eviction}\label{alg:evict}
\textbf{Input:} The expected number of predictions per eviction $k$; The TTA threshold $T$\;
  $r$ := 1\;
  \While{$r \le L$}{
   obj := Heuristic.RemoveFromTail()\;
   $TTA$ := MLModel.Predict(obj)\;
   \eIf{TTA $\ge T$}{
    break\;
   } {
    $r$ += 1\;
    Heuristic.Insert(obj, TTA)\;
   }
  }
  \If{r > k} {
   $T$ *= (1-$\delta$)\;
   }
  \If{r < k}{
   $T$ *= (1+$\delta$)\;
  }
  \textbf{return} obj\;
\end{algorithm}


As shown in Algorithm~\ref{alg:evict}, the ML thread evaluates eviction candidates one at a time by predicting its TTA and compares it with a threshold $T$. If the TTA of object $x$ is $\ge T$, object $x$ will be put on the eviction queue.  
 
If the number of iterations reaches a limit $L$, it means none of the $L$ candidates satisfies TTA $\ge T$.  In this case, we will choose the one with the largest TTA for eviction, putting it on the eviction queue.  Our system uses $L=10$ to bound the maximal cost for an eviction decision.

The threshold $T$ is dynamically adjusted to achieve a target average number of predictions per eviction, denoted as $k$.  If it takes fewer than $k$ iterations to find an object whose $TTA \ge T$, we will increase the threshold slightly $T=(1+\delta)T$. If it takes more than $k$ iterations, we will decrease the threshold slightly $T=(1-\delta)T$.  


\begin{figure}[htbp]
{
    \begin{subfigure}[t]{0.22\textwidth}
         \centering
         \includegraphics[width=\textwidth]{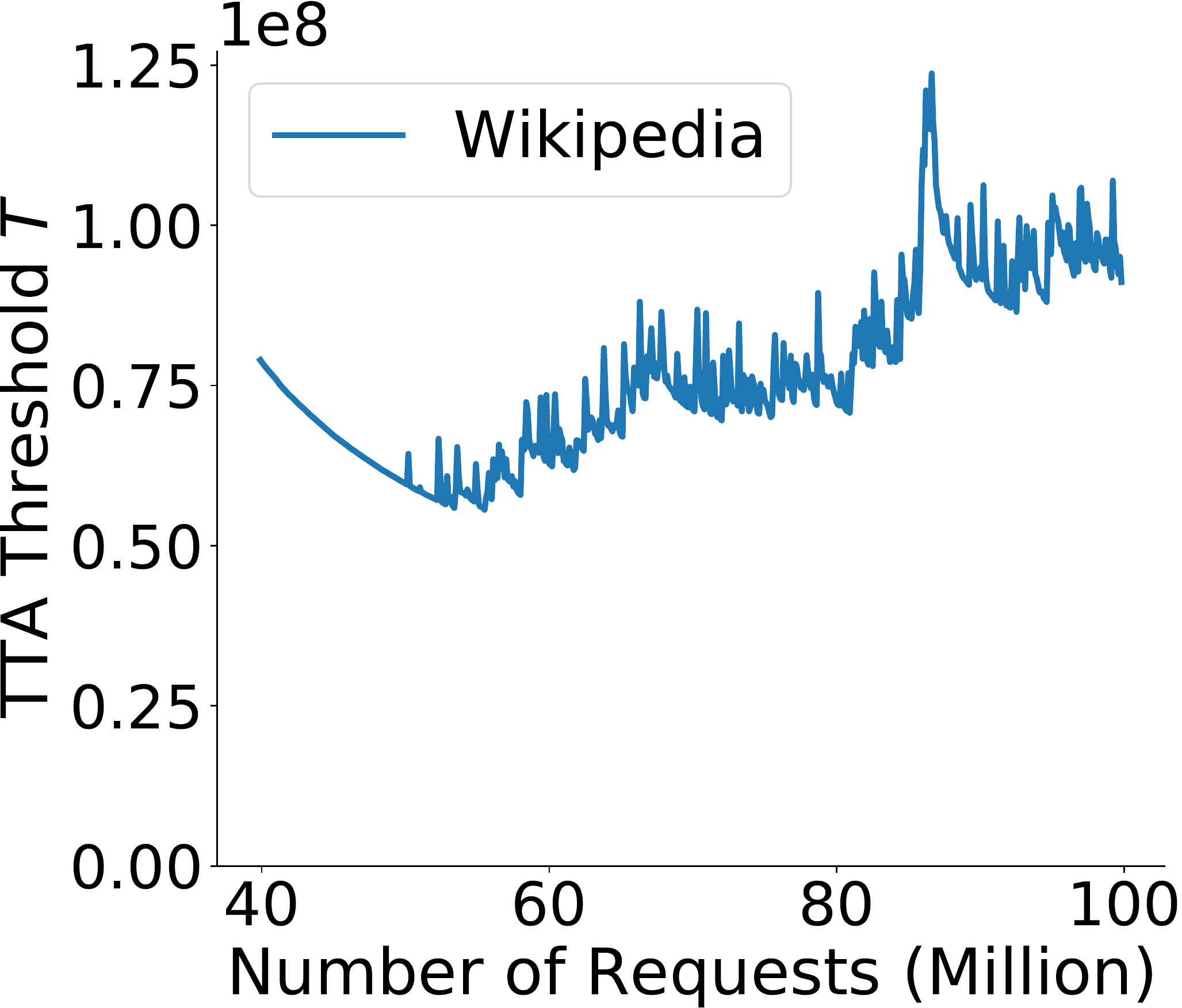}
         \caption{Optimal threshold}
         \label{fig:thr1}
     \end{subfigure}
     \begin{subfigure}[t]{0.24\textwidth}
         \centering
         \includegraphics[width=\textwidth]{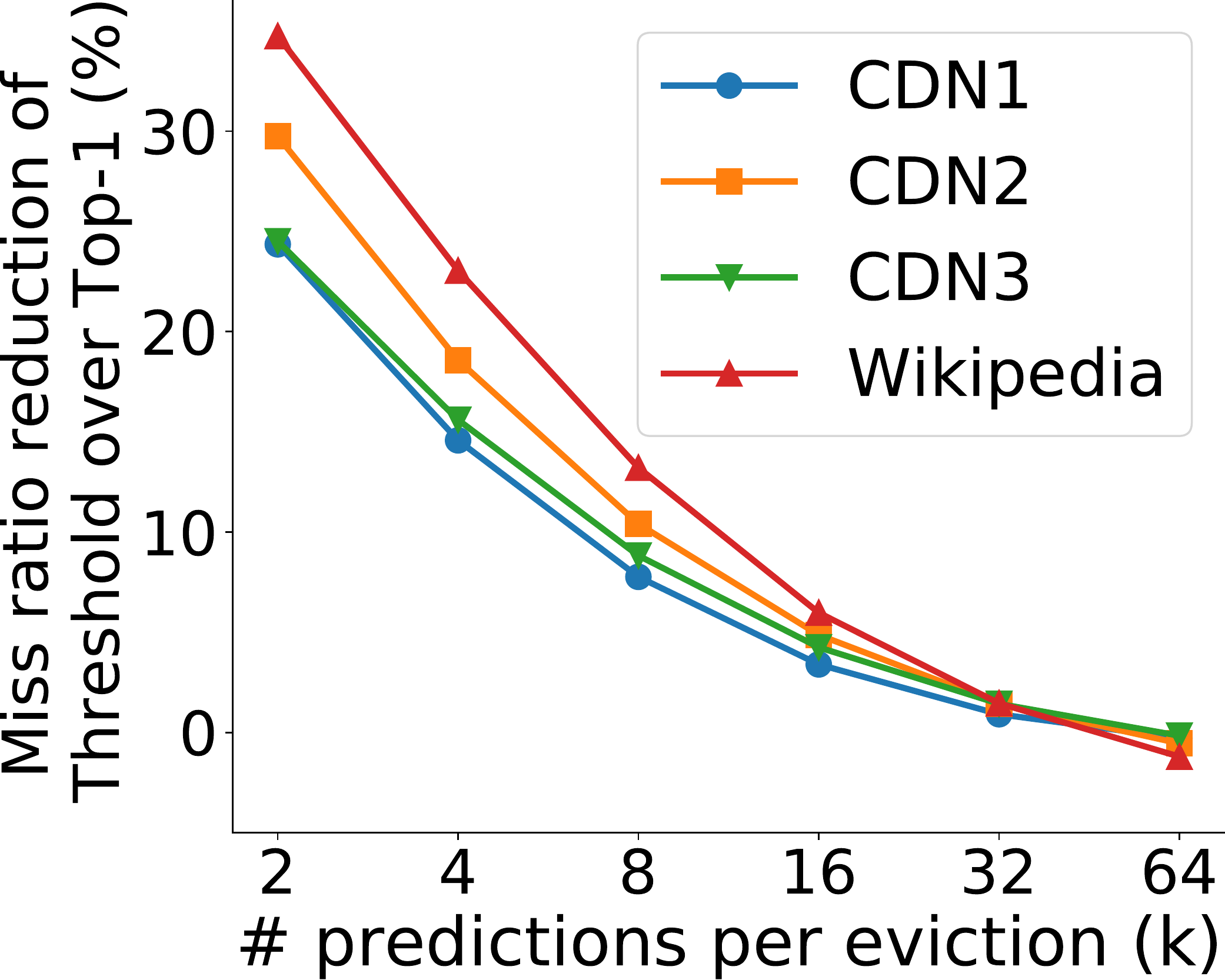}
         \caption{Threshold vs. Top-1}
         \label{fig:thr2}
     \end{subfigure}
\caption{Why using dynamic threshold to select object to evict. \textmd{(a) The best TTA threshold is continuously changing. (b) The dynamic threshold method reduces up to 30\% misses of the Top-1 method.}}
\label{fig:microbench-threshold}
}
\end{figure}


The rationale to adjust the threshold $T$ dynamically is to tolerate variations of the workloads as request distributions change over time.  We find that there is no optimal constant value of $T$ for an entire workload. 
Our system uses $\delta = 1e^{-4}$ as the default.


What happen to the objects inserted back to the priority queue?  These are the objects that the heuristic algorithm would like to evict, but the ML model disagrees.  When an object is inserted back into the priority queue, it may take a while for it to reach the tail of the queue again.  By then, its metadata (e.g., the time since last access) has changed.  The next time it becomes a candidate, the ML model might choose to evict it.

Why does \name{} uses threshold $T$ to make eviction decisions?  An alternative is to choose the object with the Top-1 TTA among $k$ eviction candidates from the heuristic algorithm.  As shown in Figure \ref{fig:microbench-threshold}, there are two advantages of using threshold $T$ over the Top-1 method.  First, in the case that the heuristic algorithm continuously sends good candidates with TTA $\ge T$, each will be selected by our threshold method, whereas the Top-1 method will miss $k-1$ good candidates.
Second, in the situation that the heuristic algorithm continuously sends bad candidates with TTA ($<T$), our threshold method choose the one with the largest TTA among $L$ candidates, whereas the Top-1 method chooses the the best one among $k$ candidates.  Since $k\ll{}L$, the Top-1 method may choose many bad candidates.

\subsection{Machine Learning Methods}
\label{sec:mlmodel}
\name{} is a general framework that can run with any supervised ML module. Here, we will introduce one ML method as an example. While this is our implementation of MAT, many other implementations can also work well.

\paragraph{Machine learning models.}
The cache request stream can be viewed as time-series data. To predict the TTA of an object, we are interested  in the past access pattern of the object  and  also in the context of other objects. The access pattern of this object can be viewed as low dimension tabular data, while the context is sequence data. 

We have experimented with several simple and efficient ML approaches including  Linear Regression (LR), Gradient Boosted Decision Trees (GBDT), Multi-layer Perceptron (MLP), and Recursive Neural Network (RNN). By default, MAT uses the GDBT model as it has a better trade-off between accuracy and computational overhead.

\paragraph{Metadata of input objects}
The ML module in the
\name{}-based caching system needs to learn from the past access patterns to perform predictions in order to make the eviction decisions to minimize cache miss ratios.  We call such data the metadata for ML and they vary depending on the choice of ML models.

To study \name{}, we use metadata similar to those in LRB~\cite{song2020learning}.
The metadata keeps track of three clusters of features for each object:
\begin{itemize}
    \item Delta$_i$ is the interval between the $i^{th}$ and the $i+1^{th}$ most recent accesses (e.g., delta$_1$ is the interval between the most recent and second most recent accesses).
    \item Exponential Decayed Counter (EDC) is a counter that is incremented on each access and is halved after a certain time. EDC$_i$ is halved after each $2^i$ requests in the cache. 
    \item  Each object also has static features that are not related to access, such as object size, object class, etc. 
\end{itemize}
By default, we maintain 32 deltas, 10 EDCs and 2 static features and the size of the metadata for each object is at most 192 Bytes. Objects with fewer past accesses will take less space.

\subsection{Training Dataset}
\label{sec:train}
In the \name{} framework, a training dataset is a set of recent historical eviction candidates, as opposed to  all objects.

The training data generation involves a tagging phase and a labelling phase. When an object is selected as an eviction candidate by the heuristic algorithm (calling \texttt{RemoveFromTail}()), it is tagged as eligible for training. When an object is requested, if it is tagged, the tag is reset and \name{} calculates the true label of TTA with regard to its last access. Then the object features and the label are inserted into the training batch. Once the training batch reaches a predefined batch size (e.g., 1 million ), it is used to retrain a ML model online and then replaces the current model.

This method uses the heuristic algorithm to filter out objects that are not heuristically determined eviction candidates.  The intuition is that they are not relevant to predictions, so excluding them will not affect the learning of the ML model. In fact, it reduces the noise in the training data and can improve the accuracy of the ML model. Depending on the miss ratio, the tagged objects can usually be 10\% to 50\% of all the objects, so the amount of the training data can be reduced by 50\% to 90\%.

\subsection{Time-To-next-Access Prediction}

The main goal of using the ML model is to predict the Time-To-next-Access (TTA) of a given object.  The inference operation with the ML model outputs the predicted distance to the next access, which is defined as the difference between the timestamps of the last access and the next access of an object. TTA is calculated as this predicted distance minus the time passed since the last access.

In the case that the predicted distance to the next access is shorter than the time passed since the last access,  we use the time passed since the last access minus the predicted distance to the next access as an estimation of TTA.

The intuition is that when the time passed since the last access is only slightly larger than the predicted distance to the next access, we still have confidence in the ML prediction and we estimate the TTA to be small. 

However, if the object still does not come and the time past since last access becomes much larger than the predicted distance to the next access, we want to stop keeping this object in the cache.

%% file: eval.tex
\section{Implementations}
\label{sec:implementation}

\subsection{Optimizations}
\paragraph{Batched predictions.}
The basic \name{} makes one eviction decision at a time. To take advantage of modern processors, \name{} can exploit the data parallelism for eviction decisions. 

This approach runs parallel predictions on $B$ objects before the tail of the heuristic algorithm. The predicted TTAs are recorded in the metadata. When the ML model receives an eviction candidate from the heuristic algorithm, it can directly use the recorded TTA for making an eviction decision.  If there is no recorded TTA, it will initiate a new parallel prediction task.

The batch size $B$ has influence on the parallelism and the miss ratio. If $B$ is too small, the parallelism is not fully realized. If $B$ is too large, the prediction results are stall and the miss ratio will be hurt. In our design, we use $B = 64$.

\subsection{Prototype}
\label{sec:prototype}
We have implemented a \name{} prototype in  Cachelib~\cite{berg2020cachelib}, which is an open-source C++ caching library.
Our implementation adds about 1,000 lines of code, with about 900 for \name{} itself and about 100 lines for integrating it into Cachelib.  We use LightGBM~\cite{ke2017lightgbm} to implement Gradient Boosted Decision Trees. 
The LightGBM model has 32 trees and each tree has no more than 32 leaves. The bagging frequency is 5 and the bagging fraction is 0.8. The learning rate is 0.1. 

\vspace{2mm}
\subsection{Simulators}
To compare \name{} to LRB, we also integrated MAT-LRU into LRB's simulation framework~\cite{lrb-code}, 
The simulator measures the miss ratios of caching algorithms by replaying all cache requests in the traces. It only maintains metadata of objects and does not allocate physical space for the objects. 

The advantages are that it can simulate cache sizes much larger than the memory on the simulation machine and the system is always bottlenecked on the caching algorithm so that we can measure the running time of the caching algorithms. 
We mainly use the simulator to perform an apple-to-apple comparison between \name{} and LRB. LRB is only available in the simulator because it is non-trivial to implement a bug-free LRB in the Cachelib prototype.
The results in Section \ref{sec:predctions} and \ref{sec:overhead} are collected from the simulator.


\section{Evaluation}
\label{sec:eval}

Our evaluation answers the following questions:
\begin{itemize}[leftmargin=*]
\vspace{-1mm}
    \item How many predictions does \name{} need for each eviction decision to achieve comparable miss ratios to SOTA? 
\vspace{-1mm}
\item What is the software overhead of \name{}?
\vspace{-1mm}
    \item What performance can \name{} prototype system achieve? 
    \vspace{-1mm}
    \item Is \name{} sensitive to heuristic algorithm choices? 
    \vspace{-1mm}
    \item How well can \name{} tolerate slow ML predictions? 
\end{itemize}
In the following, we will first describe our experimental setup, implementations, and experimental results to answer these questions.


\subsection{Experimental Setup}

Two hardware settings are used in our experiments.
All simulation experiments are run on servers in  Cloudlab~\cite{Duplyakin+:ATC19}, each with two 2\,GHz Intel E5-2683v3 CPUs (14 physical cores) and  256\,GB of RAM.

Prototype experiments are run on two servers in Microsoft Azure cloud, each with a 2.4\,GHz AMD EPYC 7763 CPU (48 cores) and 378\,GB of RAM.  The two servers are connected to 40\,Gbps local area network.



\begin{table*}[htb]
\centering
\makegapedcells
\begin{tabular}{|c|c|c|c|c|c|c|c|c|}
\hline
\multirow{2.3}{*}{\textbf{Trace}} 
& \multirow{2.3}{*}{\textbf{Type}} 
    & \multicolumn{2}{c|}{\textbf{Object Size}}
        & \multicolumn{2}{c|}{\textbf{Number of Requests}}
            & \multicolumn{2}{c|}{\textbf{Requested Bytes}} & \multirow{2.3}{*}{\textbf{\makecell{Default\\Cache Size}}}  \\
    
    \cline{3-8} & & \textbf{Mean}
        & \textbf{Max} &\textbf{Total} & \textbf{Unique}  &\textbf{Total}
        & \textbf{Unique} & \\

\hline
CDN1   & CDN  & 2\,MB   & 2\,MB    & 300\,M & 31\,M & 585\,TB & 60\,TB & 4\,TB\\
CDN2   & CDN  & 2\,MB   & 2\,MB    & 220\,M & 19\,M & 430\,TB & 38\,TB & 4\,TB\\
CDN3   & CDN  & 451\,KB   & 1\,GB    & 200\,M & 22\,M & 72\,TB & 9.5\,TB & 4\,TB\\
Wikipedia~\cite{wikipedia}   
       & CDN  & 116\,KB  & 1.3\,GB  & 200\,M & 15\,M & 7.9\,TB & 1.7\,TB & 256\,GB\\
Memcachier~\cite{memcachier} 
       & In-memory & 4.6\,KB & 1\,MB & 500\,M & 9\,M & 1\,TB & 40\,GB & 1\,GB\\
InMem & In-memory &  337\,B   & 400\,KB  & 500 M &  62\,M  & 159\,GB & 19\,GB & 8\,GB\\
IBM merged~\cite{ibm} 
      & Storage &  3.1\,M & 4\,MB  & 500\,M & 30\,M &  1832\,TB & 89\,TB & 16\,TB\\
Microsoft~\cite{microsoft} 
      & Storage &  445\,KB & 6\,MB  & 200 \,M & 48\,M & 5.1\,TB & 2\,TB & 512\,GB\\
\hline
\end{tabular}
\caption{Overview of the traces used for evaluation.
}
\label{tbl:trace}
\end{table*}

\paragraph{Workloads.}
We use 4 CDN traces and 4 other workloads in our experiments.  Table~\ref{tbl:trace} shows the characteristics of these workloads.  In addition, 
\begin{itemize}[leftmargin=*]
\vspace{-1mm}
    \item CDN1, CDN2 are collected from different caching servers (located in different regions) of same anonymous service provider. 
    \vspace{-1mm}
    \item  CDN3 is collected from the caching server of another video service provider.
        \vspace{-1mm}
    \item Wikipedia trace is from wikipedia servers.
    \vspace{-1mm}
    \item Memcachier is from a in-memory application cache.
        \vspace{-1mm}
    \item InMem is an anonymous trace collected from an in-memory key value store of social media company. 
        \vspace{-1mm}
    \item IBM merged is a combined workload of 99 traces from IBM object store, which is a cloud storage service. We merge the traces based on request timestamps.
        \vspace{-1mm}
    \item Microsoft is a storage trace from Microsoft.
\end{itemize}
 
\paragraph{Warmup.}
The first 50 million requests of each trace are used as a warm-up period. The byte miss ratios and throughput are measured after the warmup period.


We measure 3 aspects of algorithms.
\begin{itemize}[leftmargin=*]
    \item \textit{Byte Miss Ratio}:
This metric is the total size of the objects that are not present in the cache when requested divided by the total size of the objects of all requests. This metric is an indicator of the network traffic volume, which is the main optimization goal for CDN caching.
    \item \textit{Request Processing Rate}:
This metric measures the number of requests processed by the caching system per second. It is greatly influenced by the object size in the workload.
    \item \textit{Software Overhead}:
Software overhead refers to the extra computational overhead introduced by the ML algorithms. We measure the normalized running time of each part of the machine learning pipeline, including trainings, predictions, and building features. The running time is normalized by the number of evictions. In addition, we measure the number of predictions and the number of training entries incurred by the ML based caching algorithm for each eviction.
\end{itemize}

\subsection{Predictions per Eviction}
\label{sec:predctions}
\begin{figure*}[htb]
     \centering
     \begin{subfigure}[t]{0.8\textwidth}
         \centering
         \includegraphics[width=\textwidth]{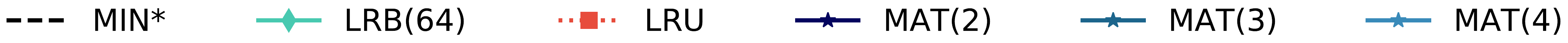}
     \end{subfigure}
         \begin{subfigure}[t]{0.245\textwidth}
         \centering
         \includegraphics[width=\textwidth]{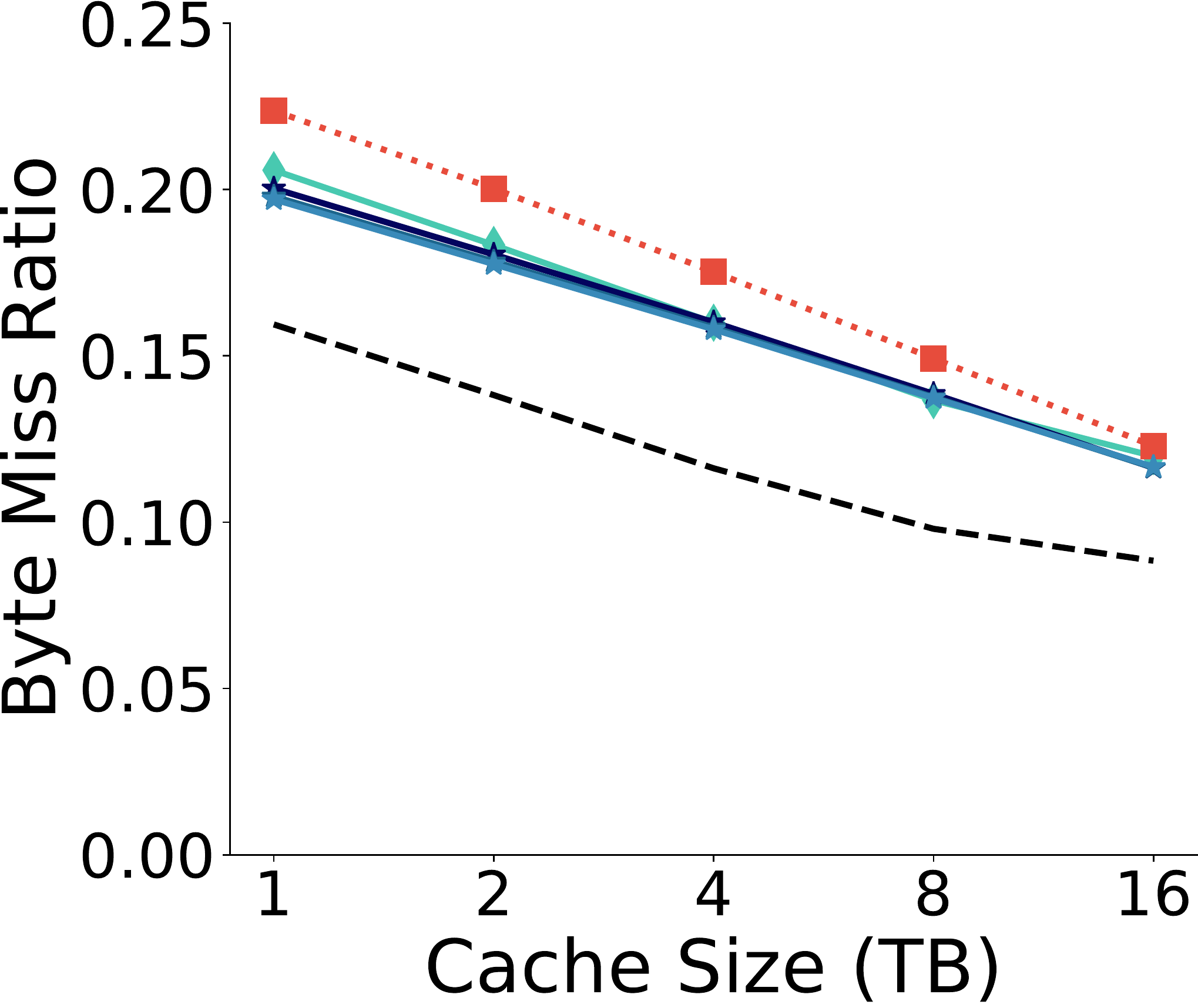}
         \caption{CDN1}
         \label{fig:s_cdn_1}
     \end{subfigure}
     \begin{subfigure}[t]{0.245\textwidth}
         \centering
         \includegraphics[width=\textwidth]{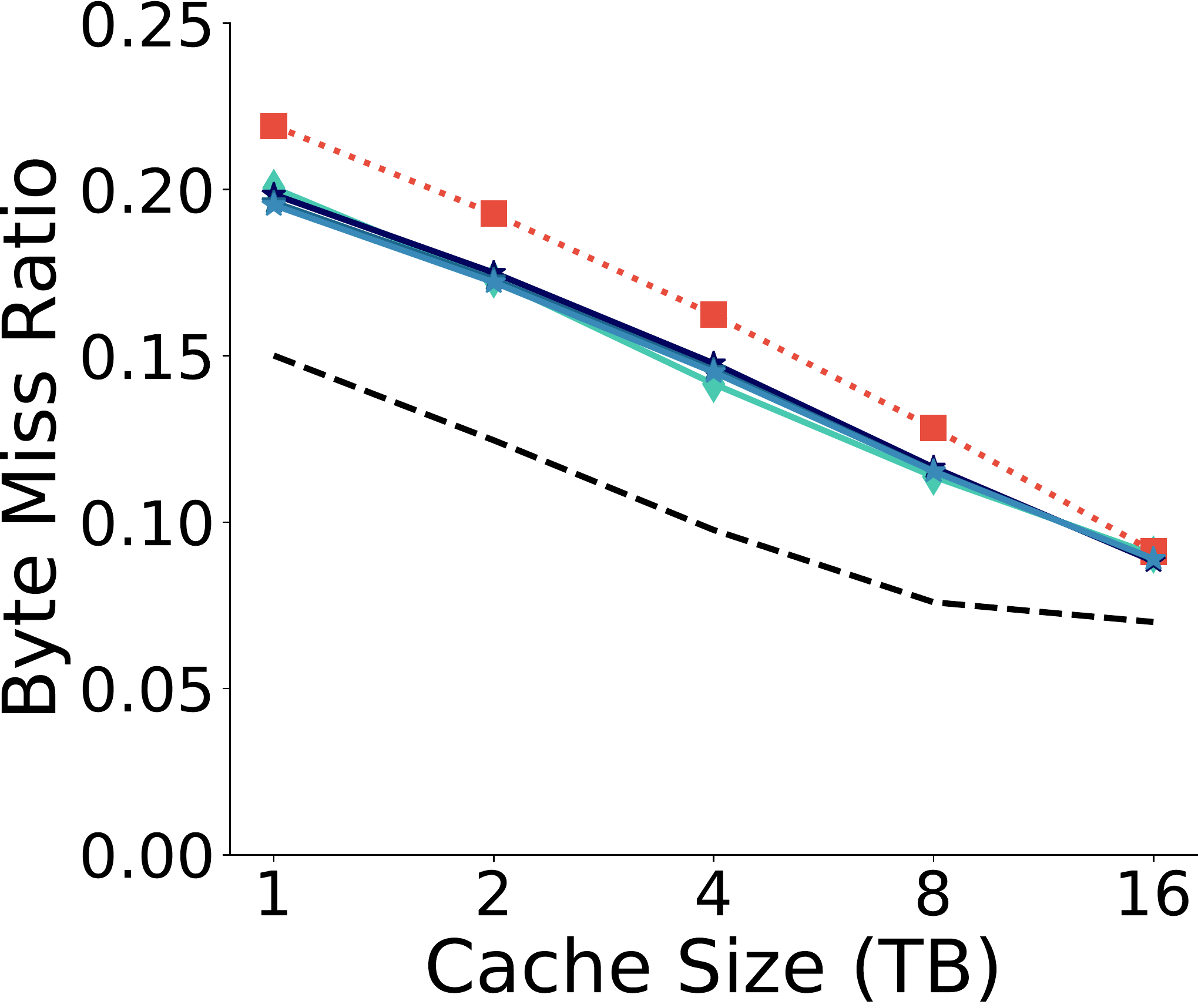}
         \caption{CDN2}
         \label{fig:s_cdn_2}
     \end{subfigure}
     \begin{subfigure}[t]{0.245\textwidth}
         \centering
         \includegraphics[width=\textwidth]{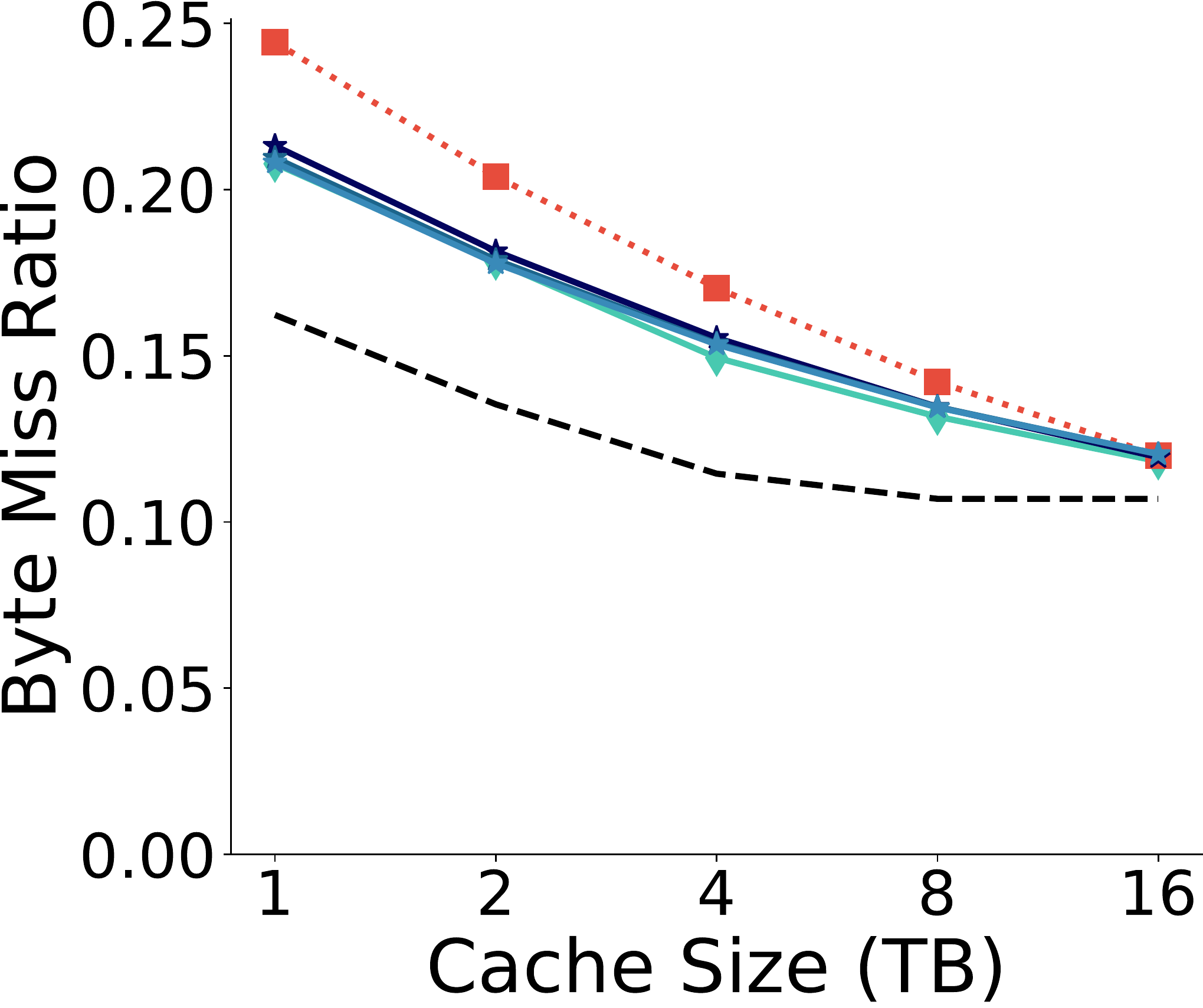}
         \caption{CDN3}
         \label{fig:s_cdn_3}
     \end{subfigure}
     \begin{subfigure}[t]{0.24\textwidth}
         \centering
         \includegraphics[width=\textwidth]{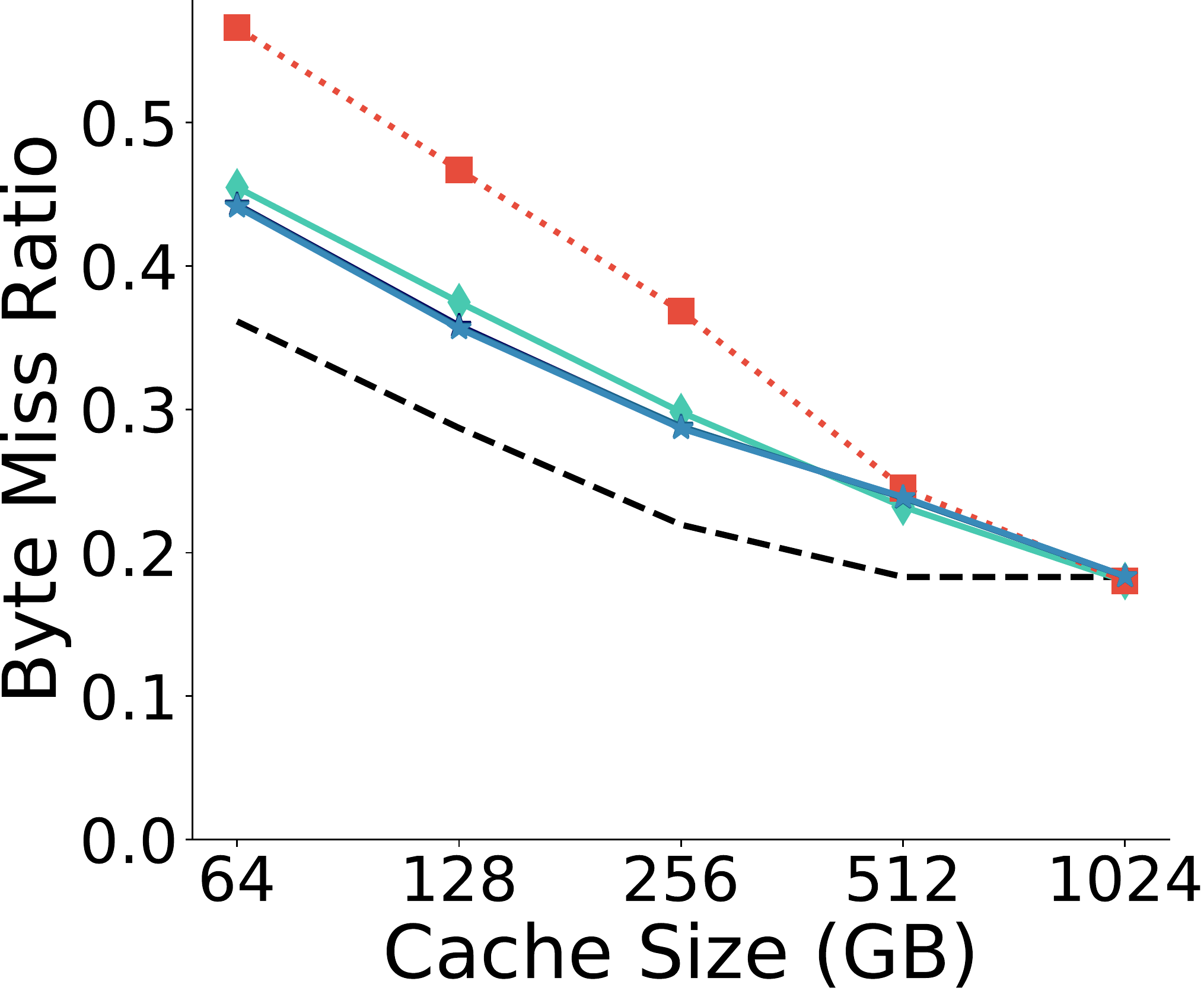}
         \caption{Wikipedia}
         \label{fig:s_cdn_4}
     \end{subfigure}
     
    \begin{subfigure}[t]{0.24\textwidth}
         \centering
         \includegraphics[width=\textwidth]{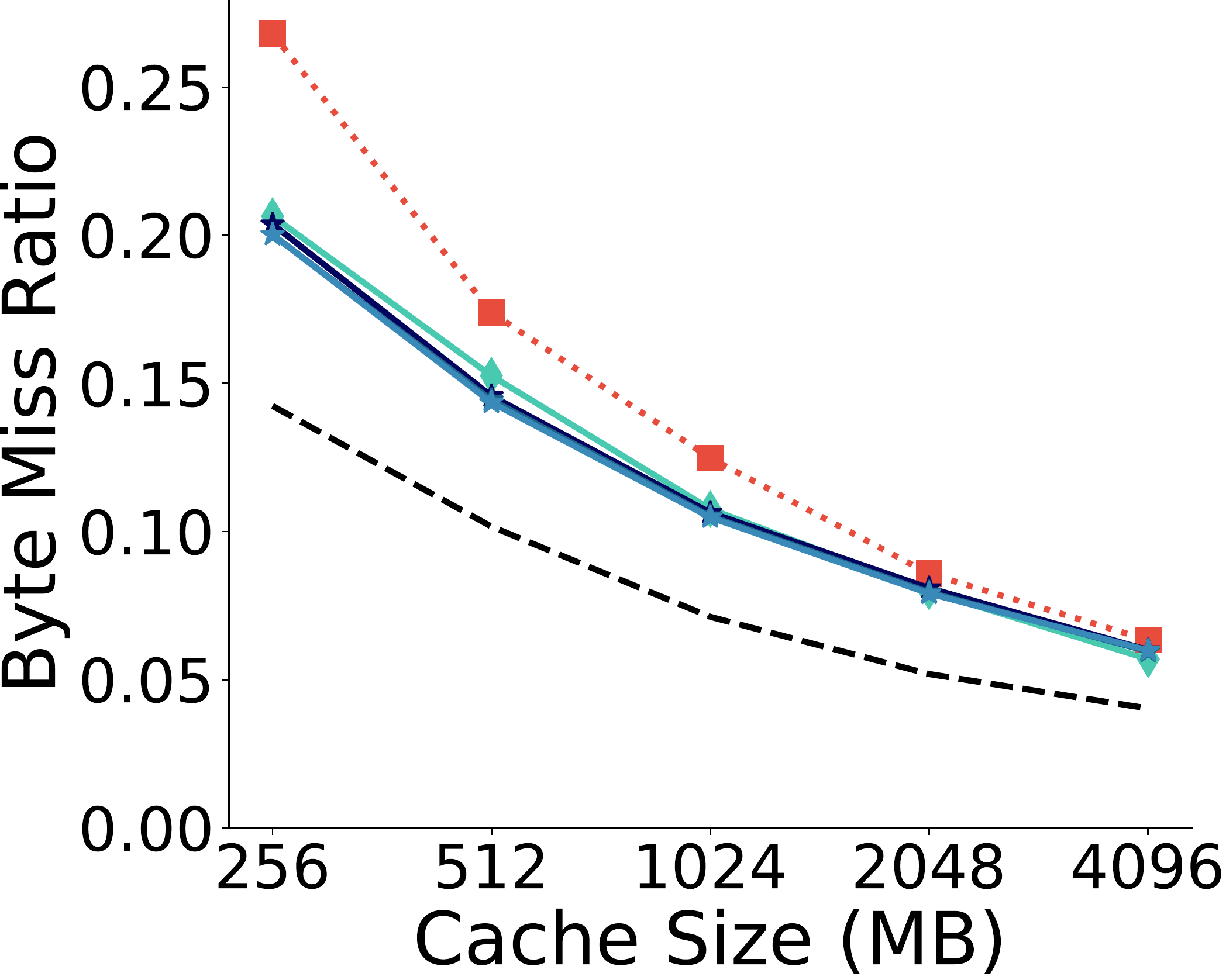}
         \caption{Memcachier}
         \label{fig:s_memc}
     \end{subfigure}
     \begin{subfigure}[t]{0.24\textwidth}
         \centering
         \includegraphics[width=\textwidth]{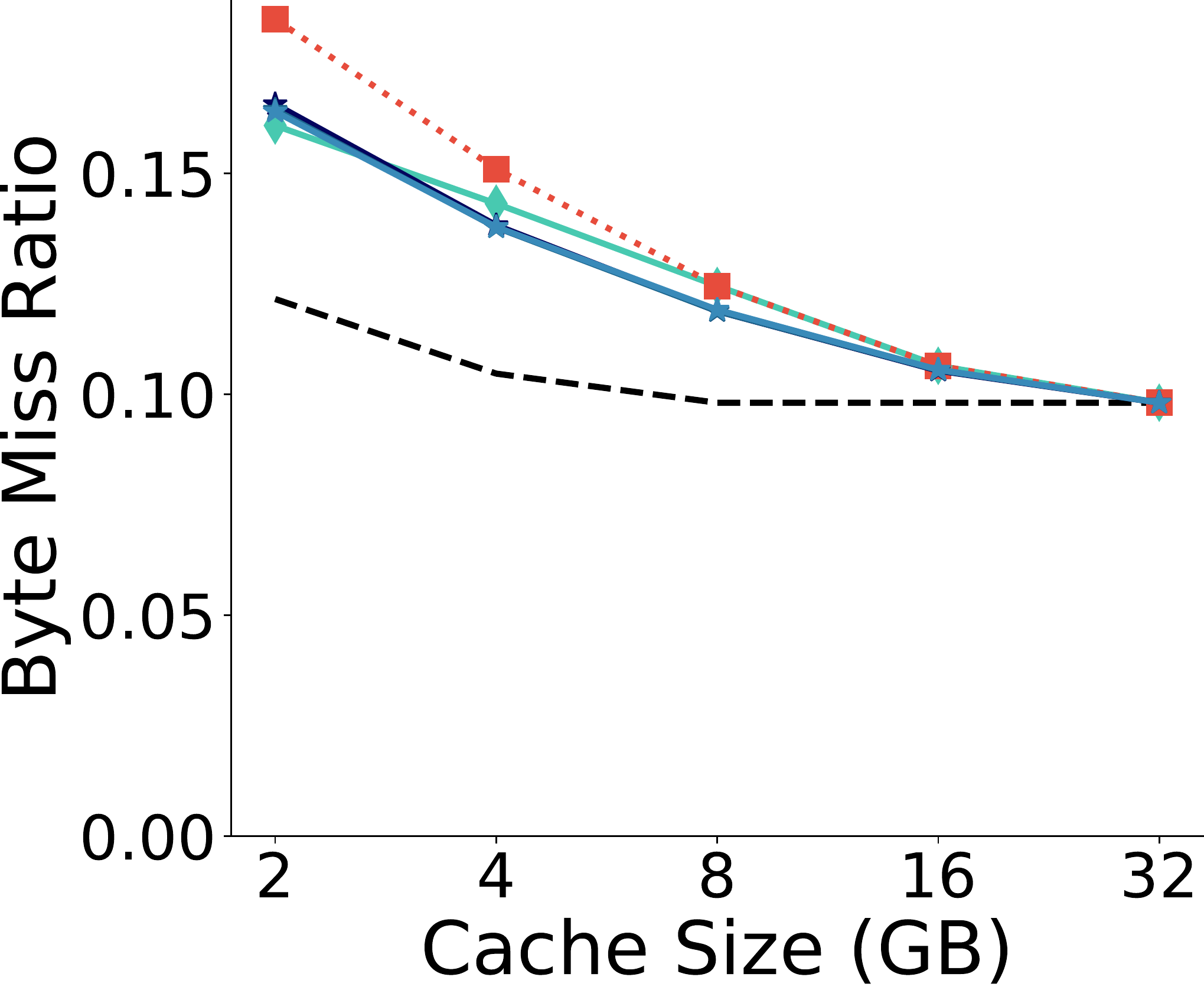}
         \caption{InMem}
         \label{fig:s_mem_1}
     \end{subfigure}
     \begin{subfigure}[t]{0.24\textwidth}
         \centering
         \includegraphics[width=\textwidth]{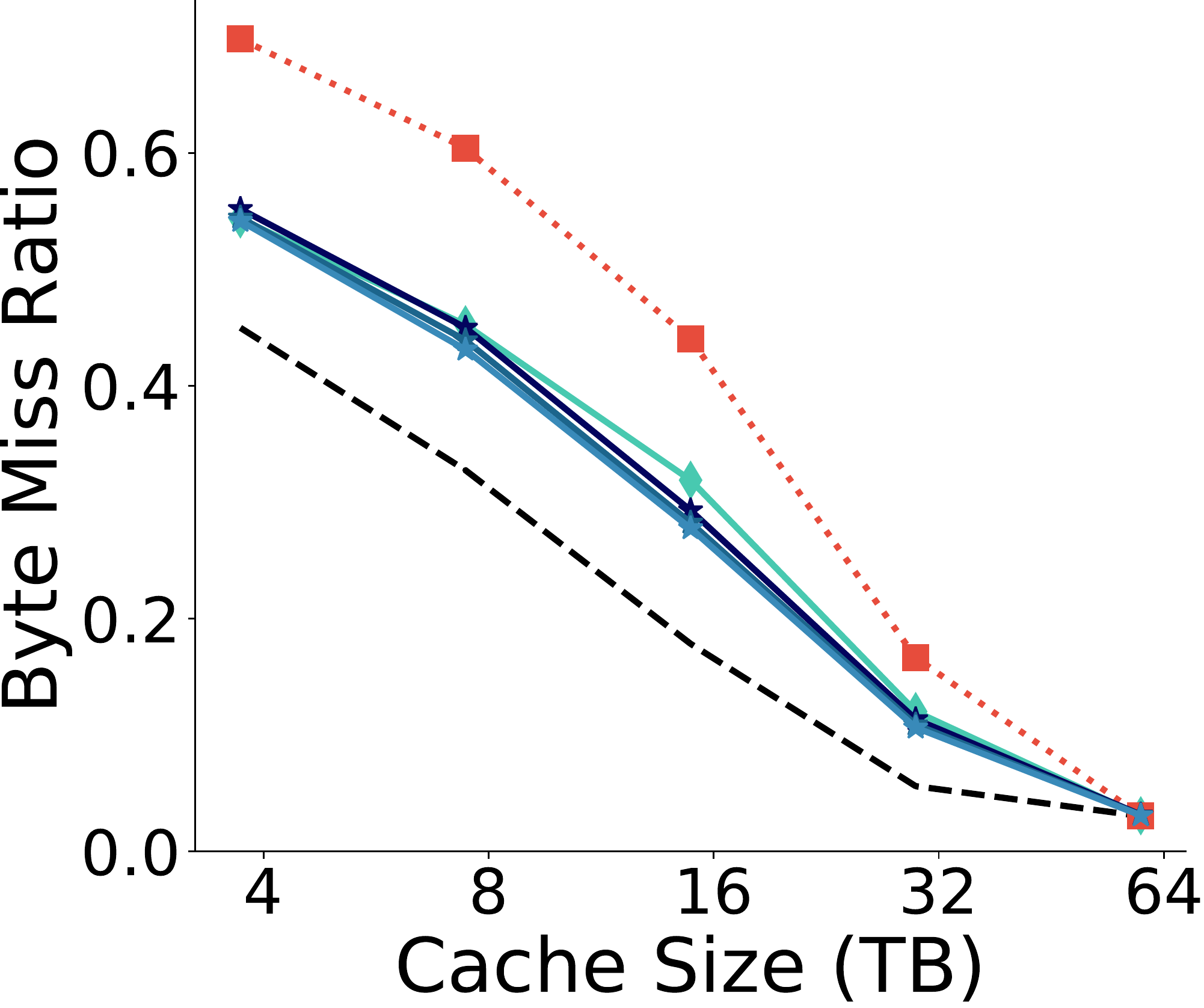}
         \caption{IBM merged}
         \label{fig:s_ibm_1}
     \end{subfigure}
     \begin{subfigure}[t]{0.24\textwidth}
         \centering
         \includegraphics[width=\textwidth]{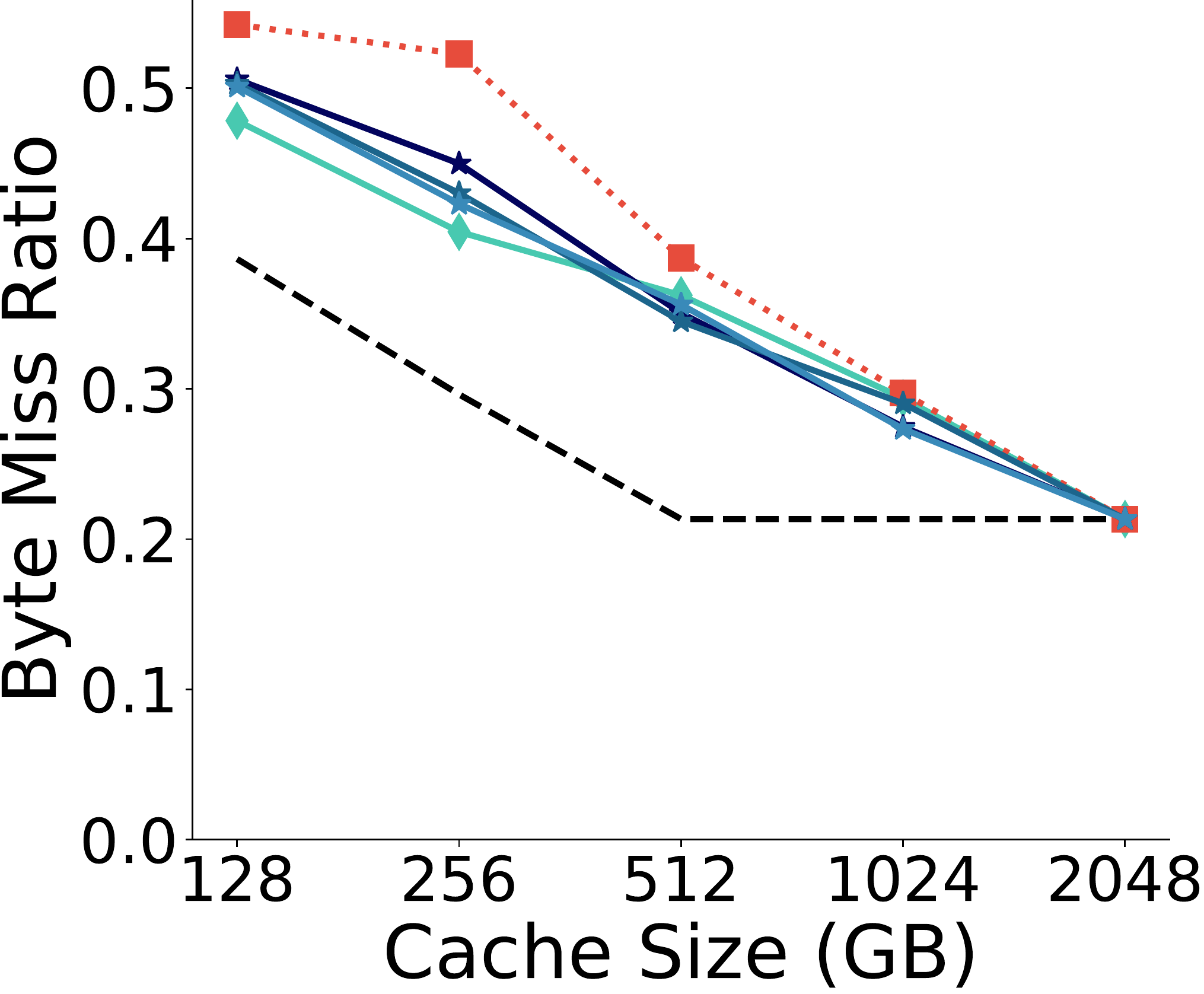}
         \caption{Microsoft}
         \label{fig:s_ibm_2}
     \end{subfigure}
\caption{Miss ratio comparisons in the simulator. \textmd{The algorithm names show the number of predictions per eviction. MAT has better or similar miss ratios compared to LRB while MAT has an order of magnitude fewer number of predictions.}
}
\label{fig:sim_miss}
\end{figure*}

To answer the first  question, we would like to experimentally evaluate how many predictions \name{} needs to make an eviction decision, while achieving similar miss ratios to the LRB~\cite{song2020learning} with various workloads.  
We conduct our experiments using the simulator with 8 workloads as shown in Figure~\ref{fig:sim_miss}. Our experiments compare 3 \name{} cases (2, 3, and 4 predictions per eviction) with LRB which runs 64 predictions per eviction (default).   


The results show that MAT reduces the number of predictions by 31 times compared to LRB without degrading the miss ratios. MAT with 2, 3, and 4 predictions per eviction have similar miss ratios to LRB with 64 predictions.

The differences among the 4 cases are small.
Figure~\ref{fig:s_cdn_1}-\ref{fig:s_cdn_4} are the results on the 4 CDN traces and the miss ratios of MAT(2), MAT(3), and MAT(4) are almost identical. MAT(2)'s miss ratios are slightly better than LRB(64)'s miss ratios on CDN1 and Wikipedia, while LRB(64) is slightly better on CDN1 and CDN2.
Figure~\ref{fig:s_memc},~\ref{fig:s_mem_1} are for in-memory traces. Compared to LRB(64), MAT(2) has 1\% and 5\% average relative reductions in the miss ratios on Memcachier and InMem, respectively.
Figure \ref{fig:s_ibm_1}, \ref{fig:s_ibm_2} show the results on storage traces. MAT(2) has in average 1\% relatively lower miss ratio than LRB(64) on IBM merged. LRB(64) has in average 2\% relatively lower miss ratio than MAT(2) on Microsoft workload.

All ML algorithms achieve significant improvements over the LRU approach.  For instance, on CDN1 with a 2\,TB cache MAT has a 18\% miss ratio compared to LRU's 21\%, which would reduce wide-area traffic by (21\%-18\%)/{21\%}=14\%.
When the cache sizes are large, the differences among them diminish. 

However, there is still a significant gap between these algorithms and the the optimal offline algorithm.  As \name{} framework can reduce the number of predictions to 2, it allows the community to explore more sophisticated ML models to reduce this gap.

\subsection{Software Overhead}
\label{sec:overhead}

To see how much overhead \name{} can reduce compared to LRB with similar ML modules, we run both in the same simulation environment with 256\,GB cache size.

Table~\ref{table:overhead} shows the average prediction overhead per eviction. The average prediction overhead including the feature building time per eviction is reduced by  reduction is 32X (from 300us to 9.3us).

\begin{table}[h]
\centering
\begin{tabular}{|l|c|c|c|}
\hline

& LRB & MAT & Reduction\\
        
\hline
Number of predictions   & 63.2   & 2.0 & 32 times \\ 
\hline

 Prediction time (us)  &  240 & 6.4 & 38 times
\\ \hline
Feature building time (us) & 60  & 2.9 & 21 times         \\\hline
Total time (us) & 300 & 9.3 & 32 times
\\ \hline
\end{tabular}
\caption{Average overhead per eviction 
(256\,GB cache size,  Wikipedia workload).}
\label{table:overhead}
\end{table}

\begin{figure*}[htbp]
     \centering
     \vspace{0.08in}
     \begin{subfigure}[t]{\textwidth}
         \centering
         \includegraphics[width=0.8\textwidth]{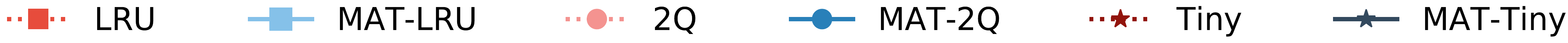}
     \end{subfigure}

     \begin{subfigure}[t]{0.243\textwidth}
         \centering
         \includegraphics[width=\textwidth]{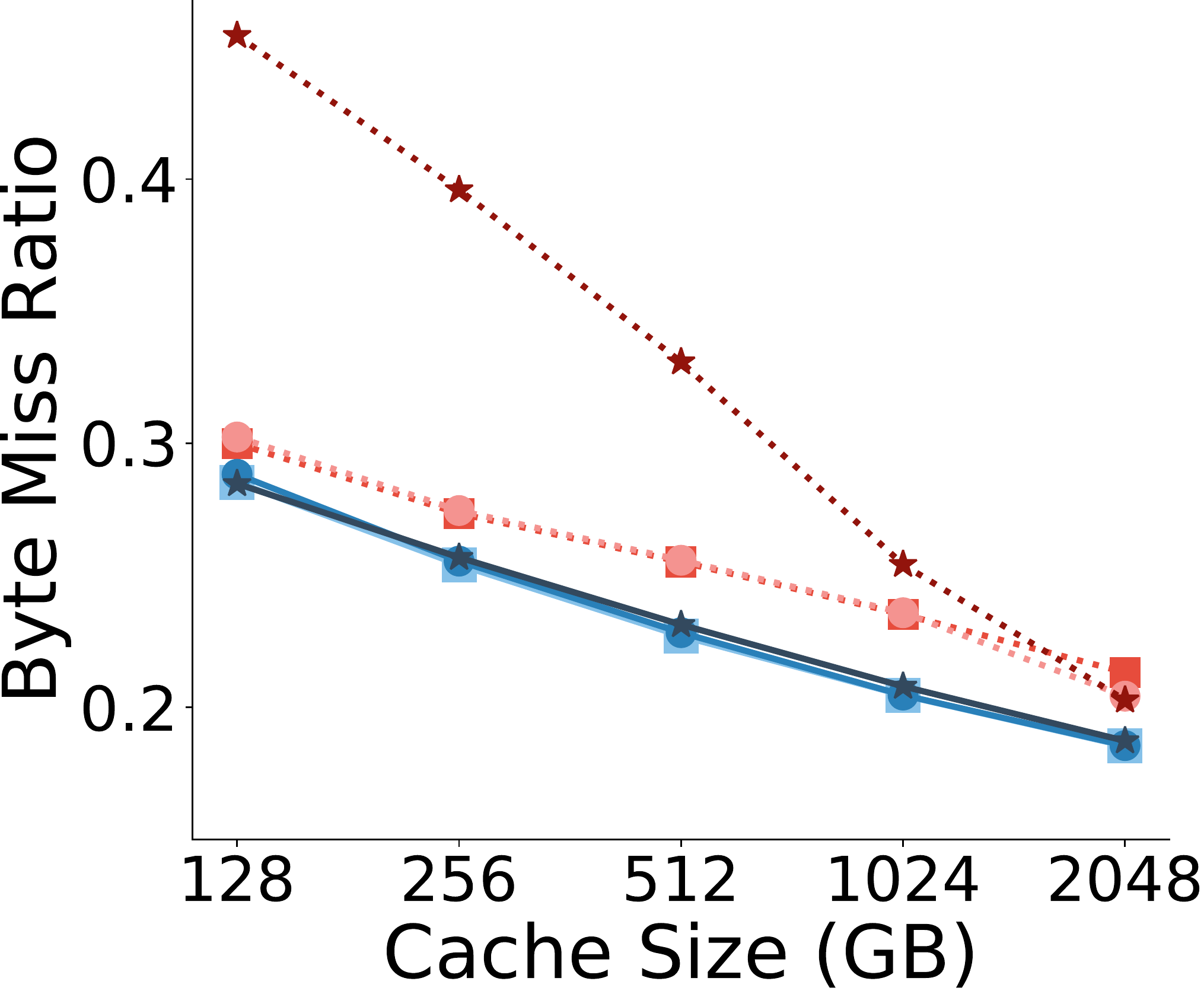}
         \caption{CDN1}
         \label{fig:b1}
     \end{subfigure}
     \begin{subfigure}[t]{0.243\textwidth}
         \centering
         \includegraphics[width=\textwidth]{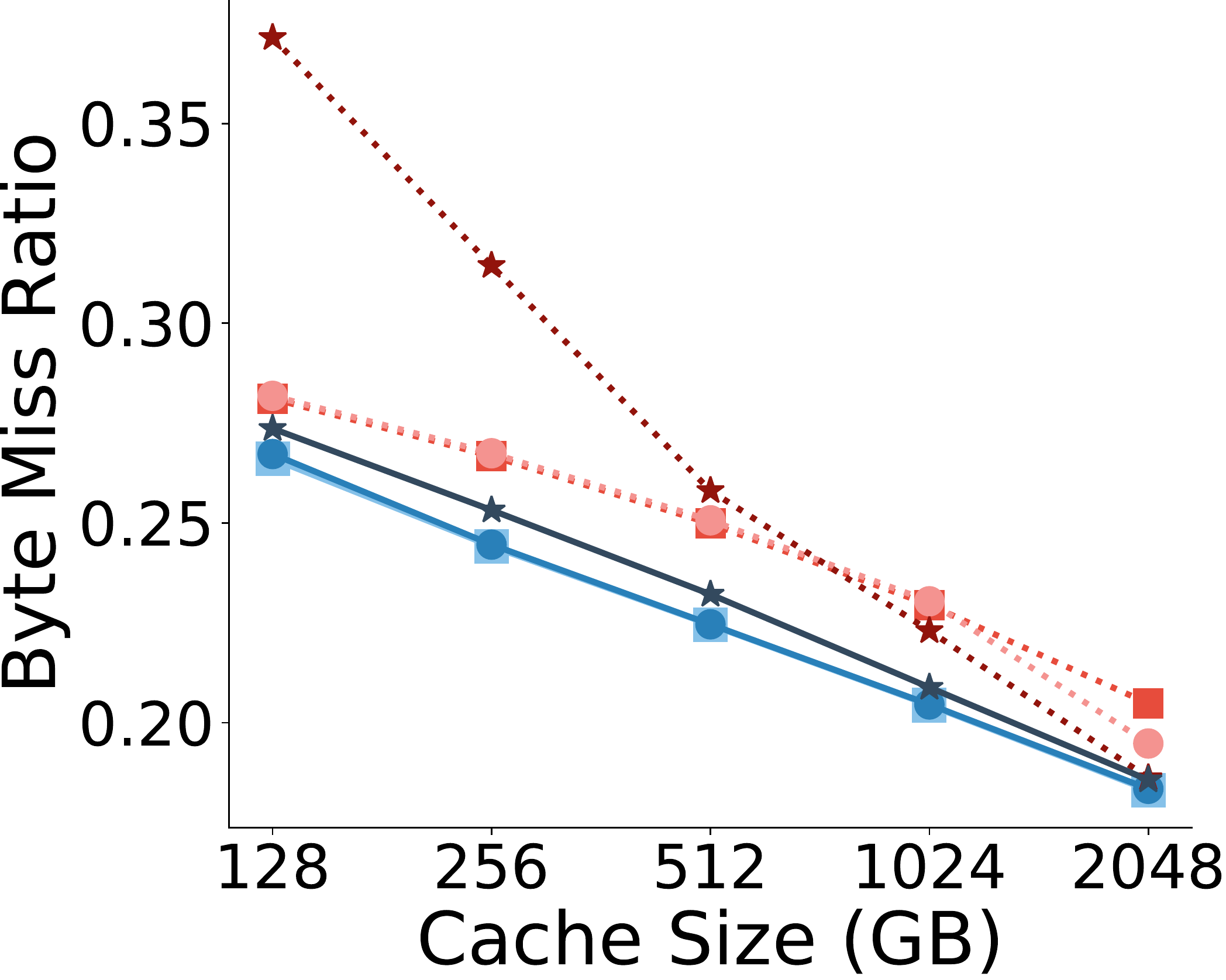}
         \caption{CDN2}
         \label{fig:b2}
     \end{subfigure}
     \begin{subfigure}[t]{0.243\textwidth}
         \centering
         \includegraphics[width=\textwidth]{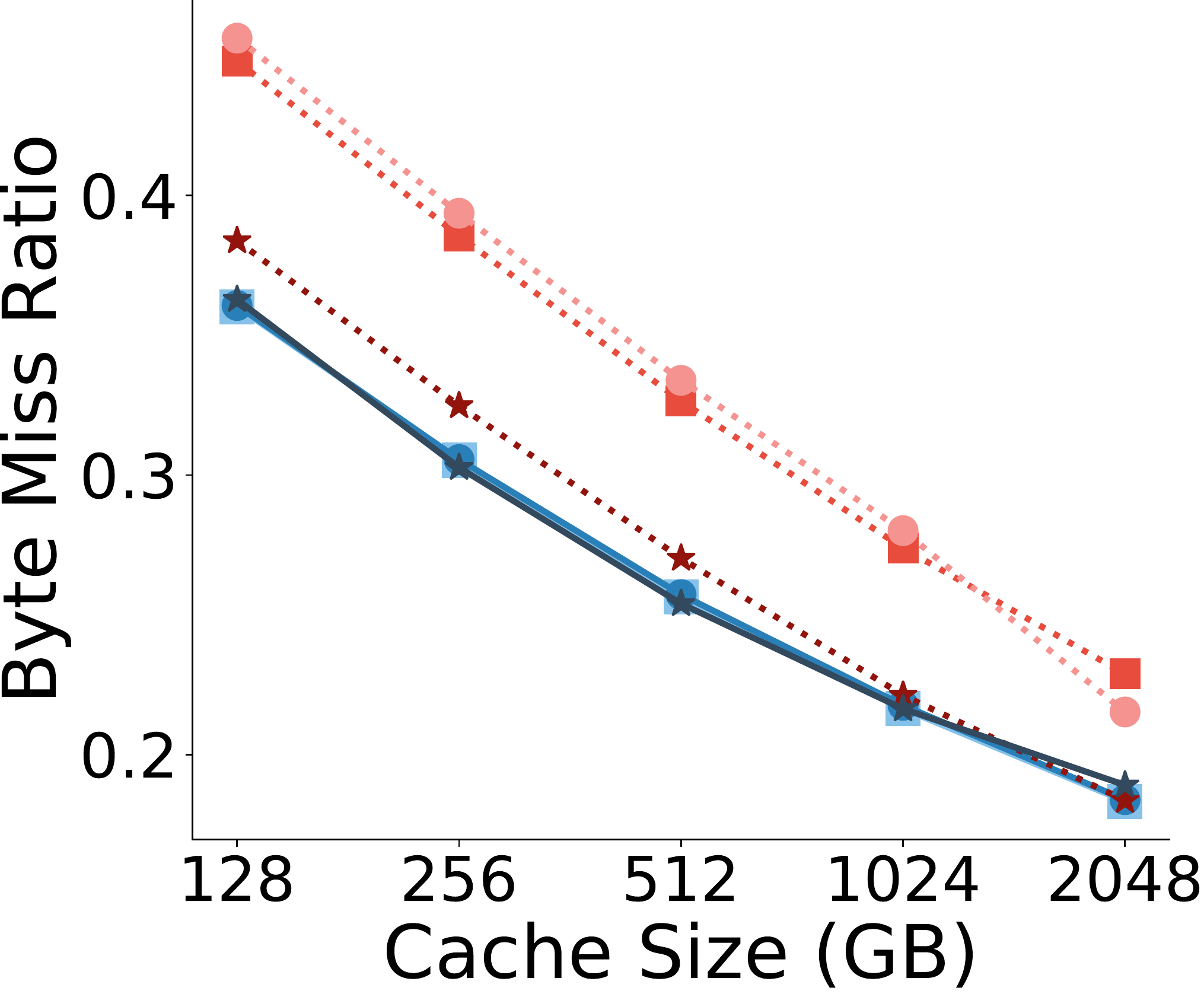}
         \caption{CDN3}
         \label{fig:b3}
     \end{subfigure}
     \begin{subfigure}[t]{0.237\textwidth}
         \centering
         \includegraphics[width=\textwidth]{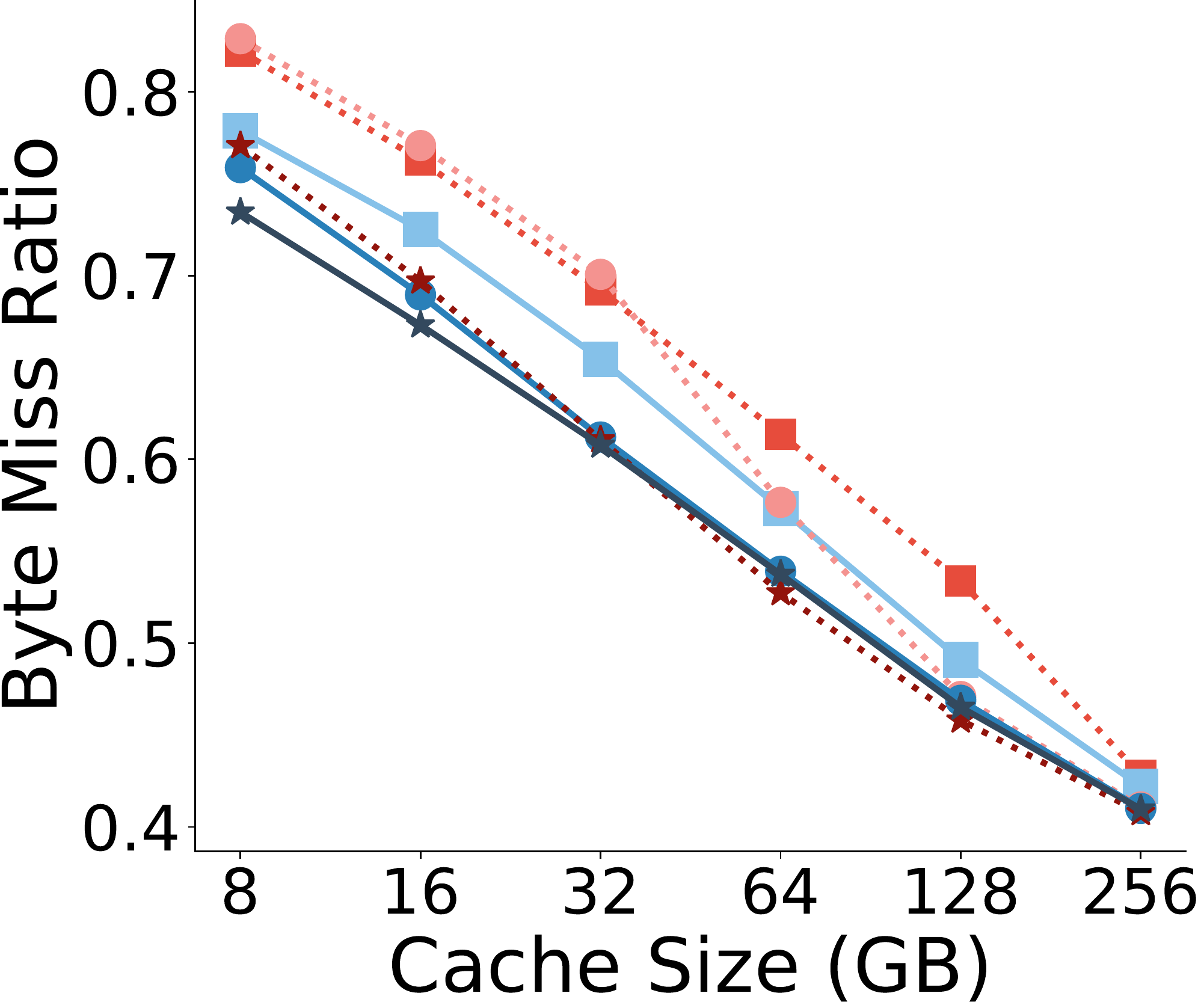}
         \caption{Wikipedia}
         \label{fig:b4}
     \end{subfigure}
\caption{The byte miss ratios of MAT with LRU, 2Q, and TinyLFU as the base algorithms. \textmd{MAT in average reduce the byte miss ratio of LRU, 2Q, and TinyLFU by relatively 12\%, 12\%, and 10\% respectively.}}
\label{fig:base_miss}
\end{figure*}

MAT reduces the average number of predictions by 31X (from 63.2 to 2) while reducing the average feature building time by 21X (from 60us to 2.9us).  
Feature building refers to converting the object metadata to the feature matrix which serves as the input of the ML model, for both training and predictions.


Table~\ref{table:overhead_training} shows the average overheads per eviction of LRB and MAT.
The average training time reduction of MAT is 9.5X
(from 160us to 16.9us).

\begin{table}[h]
\centering
\begin{tabular}{|l|l|l|l|l|}
\hline
                           & LRU & MAT & Reduction \\ \hline
Number of training samples & 8.6 & 0.93 &  9 times     \\ \hline
Training time (us)         & 100 & 14 & 7 times   \\ \hline

\end{tabular}

\caption{Average training overhead per eviction ( 256\,GB cache size, Wikipedia workload).}
\label{table:overhead_training}
\end{table}

The average number of training samples per eviction is reduced by 9.2X (from 8.6 to 0.93).  The average training time overhead is reduced by 7X (from 100us to 14us).



In summary, the total overhead of MAT's ML module is 23.3\,us per eviction, 17X reduction over LRB.  

\subsection{Prototype Performance}

To evaluate the performance of \name{} prototype, we compare its performance with Cachelib with LRU algorithm.  We are interested in the request processing rate of \name{} prototype compared to that of LRU-based caching system.  Since both \name{} and LRU are implemented in Cachelib.  We conduct experiments in the same setting.

To run such experiments, we extended the Cachebench module in the Cachelib library to support running experiments over a network. We implement a Cachebench client instance, a Cachebench server instance, and an Nginx server instance.  The Cachebench client instance reads requests from a trace file and sends them through HTTP to the Nginx server using CURL. The Nginx server  accepts requests and forwards them to the Cachebench server instance using Fastcgi. The Cachebench server runs either prototype which executes the requests and returns the results using the Fastcgi API to the Nginx server.  The Nginx server then forwards the requests back to the Cachebench client. 

\begin{table}[h]
\centering
\begin{tabular}{|l|l|l|l|}
\hline
Trace                      & Cache Size & MAT         & LRU         \\ \hline
\multirow{3}{*}{Wikipedia} & 32\,GB       & 23787 req/s & 24465 req/s \\ \cline{2-4} 
                           & 128\,GB      & 25117 req/s & 25577 req/s \\ \cline{2-4} 
                           & 512\,GB      & 30258 req/s & 31181 req/s \\ \hline
Memcachier                 & Infinite   & 52883 req/s & 51380 req/s \\ \hline
\end{tabular}
\caption{Request rates of MAT and LRU prototypes with Wikipedia and Memcachier workloads.}
\label{tab:prototype}
\end{table}

The main result is that \name{} prototype achieves similar request rates as LRU prototype.  Table~\ref{tab:prototype} shows the request rates of MAT and LRU with Wikipedia and Memcacheir workloads.

For Wikipedia workload whose average object size is 116KB, \name{} achieves 23,878, 25,117, and 30,258 requests/sec with cache sizes of 32GB, 128GB, and 512GB respectively. It is 2.8\%, 1.8\%, and 3.0\% slower than LRU.
The these results include the warmup period.  Without the warmup, we expect \name{} will achieve higher request rates than above since its miss ratio is substantially lower than LRU.

The reason for using Memcachier workload is to test maximal request rates as its average object size is relatively small (4.6KB).  \name{} achieves 52,883 requests/sec whereas LRU achieves 51,380 requests/sec.  In this case, \name{} is 2.9\% faster.



\vspace{0.05in}
\subsection{Heuristic Algorithm Choices}
\label{sec:heuristic}

To understand the effects of different heuristic algorithm choices, we compare three heuristic algorithms (LRU, 2Q and TinyLFU) in Cachelib with their corresponding \name{} implementations (MAT-LRU, MAT-2Q and MAT-TinyLFU) as shown in 
Figure~\ref{fig:base_miss}.  We conduct experiments with 
4 workloads: CDN1, CDN2, CDN3 and Wikipedia.

The experiments show two results.  First,  \name{} is not sensitive to heuristic algorithm choices.  MAT-LRU, MAT-2Q and MAT-TinyLFU  achieve similar miss ratios with all 4 workloads.  MAT-TinyLFU achieves slightly worse miss ratios than MAT-LRU and MAT-2Q with CDN2 workload.  

Second, \name{} can correct the eviction mistakes by its heuristic algorithm.  Dramatic examples are TinyLFU with CDN1 and CDN2.  In both cases, TinyLFU has much higher miss ratios than LRU and 2Q.  However, MAT-TinyLFU can achieve miss ratios similar to those of MAT-LRU and MAT-2Q. The ML model in \name{}  can effectively make good eviction decisions, adapting to different request patterns.

\vspace{0.09in}
\subsection{Tolerance to Slow ML Predictions}
\label{sec:slow}

\begin{figure}[htbp]
         \centering
         \includegraphics[width=0.3\textwidth]{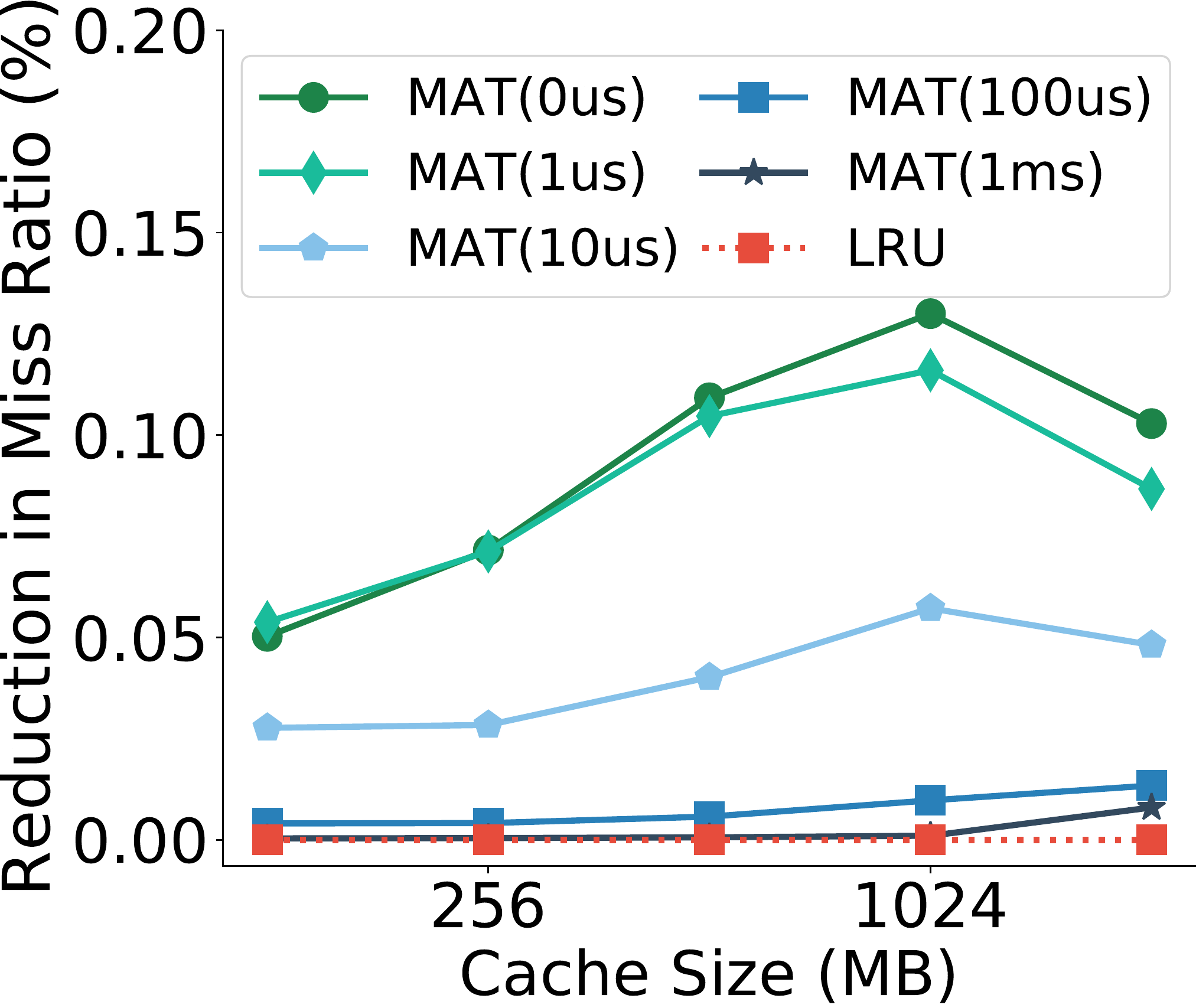} 
\caption{MAT with slow Predictions. \textmd{When the ML model is slower, the miss ratio degenerates gracefully.}}
\label{fig:slowdown}
\end{figure}

\name{} has the ability to fall back to its heuristic algorithm when its ML threads cannot keep up with the rate of evictions.
To answer the question how well \name{} can tolerate slow ML predictions, we conduct experiments on CDN1 by stalling each prediction for 1\,us, 10\,us, 100\,us, and 1\,ms. Note that a prediction itself takes 3\,us on average.


Figure \ref{fig:slowdown} shows that the miss ratio reduction of MAT using LRU as the baseline. MAT can always deliver better or equal miss ratio to LRU and MAT degrades gracefully when the ML model stalls. When the stall is 10\,us, which means the ML model runs at 25\% of its full speed, the miss ratio reduction is about 40\% of the full speed ML model. This makes MAT particularly practical for deployment because it can run elastically with any available amount of computation resources.




%% file: related.tex
\section{Related Work}
\label{sec:related}
This section discusses related work on learning-based cache algorithms and heuristic cache algorithms. 
We also review systems targeting other aspects of CDN cache system.

\paragraph{Learning-based cache replacement}


Existing works on applying machine learning cache algorithms can be categorized into supervised learning and reinforcement learning.
Most supervised learning approaches use regression.
LRB~\cite{song2020learning} is the most similar approach to MAT.
As we have seen in Section~\ref{sec:eval}, LRB processing time overhead is $17\times$ higher than MAT.
LFO~\cite{berger2018towards} uses boosting tress to predict and imitate the admission policy of OPT. 
It requires complex offline training, and thus is not practical to adapt quickly to workload changes.
LFO also performed poorly in prior experiments~\cite{song2020learning}.
Similar to regression on a single value, \cite{zhou2021learning} learns the next access distribution, and uses it to compute a utility. 
However, to get the distribution training data, it is also limited to offline training.
Different from regression, Parrot~\cite{liu2020imitation} takes a ranking approach. Instead of predicting the time to next access, it directly learns to rank objects. 
This approach is more end-to-end but the computation overhead is much higher than regression.

Another line of works~\cite{wang2019learning, kirilin2020rl, costa2017mlcache} apply reinforcement learning to cache algorithms.
Instead of predicting the time to next access, the eviction policy is directly modeled and optimized.
Unfortunately, feedback loops in production systems would be in the tens of millions of steps, which exceeds the capability of current reinforcement learning frameworks.
Consequently, the performance of reinforcement-learning-based approaches is worse than supervised learning.

MAT is the first learning-based algorithm that can provide both a low miss ratio and high throughput. This is because its design provides efficient online training and inference.

\paragraph{Heuristic-based cache replacement}
Over the past five decades, many cache algorithms based on heuristics have been proposed. 
Prominent examples include LRU, FIFO, SLRU~\cite{karedla1994caching}, 2Q~\cite{johnson19942q}, LIRS~\cite{jiang2002lirs}, ARC~\cite{megiddo2003arc}, LeCaR~\cite{vietri2018lecar}, and LHD~\cite{beckmann:nsdi18:lhd}.
MAT is agnostic to most base heuristic algorithms and can use any priority-queue based heuristic as its base algorithm.



\paragraph{Learning-augmented cache replacement}

Several learning-augmented replacement algorithms have been proposed to combine heuristic with learning-based algorithms. To the best of our knowledge, previous works~\cite{lykouris2018competitive,rohatgi2020near,wei2020better,antoniadis2020online} focus on theoretical analysis on the competitive ratio of learning-augmented cache, or only evaluate the cache miss ratio regardless of the overhead~\cite{chlkedowski2021robust}. MAT is the first learning-augmented cache that achieves a low byte miss ratio and high throughput on production cache software and workloads.

\paragraph{Cache admission policies}
Multiple systems focus on learning smart cache admission policies while relying on existing cache replacement heuristics such as TinyLFU~\cite{einziger2014tinylfu, einziger2018adaptive}, AdaptSize~\cite{berger2017adaptsize}, Flashield~\cite{eisenman2019flashield}, and CacheLib~\cite{berg2020cachelib}.
While there is no public implementation of Flashield, our evaluation of TinyLFU, AdaptSize, and CacheLib shows that admission policies are not sufficient to achieve state-of-the-art miss ratios.
LRB outperforms both systems' miss ratio, and MAT achieves similar miss ratios to LRB while significantly improving throughput and reducing overhead.

\paragraph{High-throughput in-memory caching systems}
MemC3~\cite{fan2013memc3} and LHD~\cite{beckmann:nsdi18:lhd} show how to significantly increase the throughput of in-memory caching systems based on replacement heuristics.
The throughput challenges faced by a ML-based caching systems are different from the setting in MemC3.
Some of MemC3's techniques, such as faster hashing and fast approximations for a heuristic like LRU, are complementary to MAT. In fact, MemC3's replacement heuristic can be used to create candidates for MAT's eviction filter.
LHD's primary design based on sampling is similar to LRB.
MAT effectively overcomes the challenges of sampling which requires too many evaluations and calls to a prediction model.
SegCache \cite{yang2021segcache} is designed for small objects with TTLs. It groups objects with similar TTLs together to reduce memory fragmentation and thus improves the hit ratios.

\paragraph{CDN cache systems}
Many works optimize CDN cache systems from other aspects. 
RIPQ~\cite{tang2015ripq} co-locates small writes to reduce SSD write amplification. 
AViC~\cite{akhtar2019avic} designs the eviction algorithm based on the video chunk sequential accessed at a constant speed and leverages the properties of video delivery to optimize the hit ratio.
The design of MAT is flexible for general cache and can be applied together with these systems.

%% file: concl.tex
\section{Conclusion}
\label{sec:concl}

MAT is proposed as a general framework for building an efficient ML-based caching system by adding an ML module to an existing cache system based on a heuristic algorithm. 

The key idea behind MAT is to treat a heuristic algorithm as a filter for the ML module.  Most heuristic algorithms can serve as good filters, as we demonstrate that they evict most objects an optimal algorithm evicts while evicting some objects they should keep.  The role of ML module is to correct these mistakes.

We show several simulation and prototyping results.  
First, it reduces the average number of predictions per eviction to 2, a 31 times reduction compared to the state-of-the-art ML-based caching system while achieving comparable miss ratios. Second, MAT is not sensitive to the choice of heuristic algorithms.  Third, MAT can  fall back to the heuristic algorithm which allows it to 
run efficiently, tolerating slow ML inferences or lack of computing power.


The ML module used in our implementations is Gradient Boosted Decision Tree (GBDT) due to its simplicity and efficiency.  Other ML methods that are less efficient may further reduce miss ratios.   Since MAT framework dramatically reduces the ML overhead to merely 2 predictions per eviction, it enables us to explore some sophisticated ML approaches.





%% file: main.bbl
\begin{thebibliography}{44}
\providecommand{\natexlab}[1]{#1}
\providecommand{\url}[1]{\texttt{#1}}
\expandafter\ifx\csname urlstyle\endcsname\relax
  \providecommand{\doi}[1]{doi: #1}\else
  \providecommand{\doi}{doi: \begingroup \urlstyle{rm}\Url}\fi

\bibitem[ibm()]{ibm}
Ibm object storage.
\newblock \url{http://iotta.snia.org/traces/key-value/36305}.

\bibitem[lrb()]{lrb-code}
Lrb github repository.
\newblock \url{https://github.com/sunnyszy/lrb}.

\bibitem[mem({\natexlab{a}})]{memcached}
memcached - a distributed memory object caching system.
\newblock \url{http://memcached.org/}, {\natexlab{a}}.

\bibitem[mem({\natexlab{b}})]{memcachier}
Memcachier.
\newblock \url{https://www.memcachier.com/}, {\natexlab{b}}.

\bibitem[mic()]{microsoft}
Microsoft.
\newblock \url{http://iotta.snia.org/traces/block-io/388}.

\bibitem[red()]{redis}
redis.
\newblock \url{https://redis.io/}.

\bibitem[wik()]{wikipedia}
Wikipedia trace.
\newblock \url{http://lrb.cs.princeton.edu/wiki2019.tr.tar.gz}.

\bibitem[Akhtar et~al.(2019)Akhtar, Li, Govindan, Halepovic, Hao, Liu, and
  Sen]{akhtar2019avic}
Z.~Akhtar, Y.~Li, R.~Govindan, E.~Halepovic, S.~Hao, Y.~Liu, and S.~Sen.
\newblock Avic: a cache for adaptive bitrate video.
\newblock In \emph{Proceedings of the 15th International Conference on Emerging
  Networking Experiments And Technologies}, pages 305--317, 2019.

\bibitem[Antoniadis et~al.(2020)Antoniadis, Coester, Elias, Polak, and
  Simon]{antoniadis2020online}
A.~Antoniadis, C.~Coester, M.~Elias, A.~Polak, and B.~Simon.
\newblock Online metric algorithms with untrusted predictions.
\newblock In \emph{International Conference on Machine Learning}, pages
  345--355. PMLR, 2020.

\bibitem[Beckmann et~al.(2018)Beckmann, Chen, and Cidon]{beckmann:nsdi18:lhd}
N.~Beckmann, H.~Chen, and A.~Cidon.
\newblock {LHD}: Improving cache hit rate by maximizing hit density.
\newblock In \emph{USENIX NSDI}, pages 389--403, 2018.

\bibitem[Belady(1966)]{belady}
L.~A. Belady.
\newblock A study of replacement algorithms for a virtual-storage computer.
\newblock \emph{IBM Systems journal}, 5\penalty0 (2):\penalty0 78--101, 1966.

\bibitem[Berg et~al.(2020)Berg, Berger, McAllister, Grosof, Gunasekar, Lu,
  Uhlar, Carrig, Beckmann, Harchol-Balter, et~al.]{berg2020cachelib}
B.~Berg, D.~S. Berger, S.~McAllister, I.~Grosof, S.~Gunasekar, J.~Lu, M.~Uhlar,
  J.~Carrig, N.~Beckmann, M.~Harchol-Balter, et~al.
\newblock The cachelib caching engine: Design and experiences at scale.
\newblock In \emph{14th $\{$USENIX$\}$ Symposium on Operating Systems Design
  and Implementation ($\{$OSDI$\}$ 20)}, pages 753--768, 2020.

\bibitem[Berger(2018)]{berger2018towards}
D.~S. Berger.
\newblock Towards lightweight and robust machine learning for cdn caching.
\newblock In \emph{ACM HotNets}, pages 134--140, 2018.

\bibitem[Berger et~al.(2017)Berger, Sitaraman, and
  Harchol-Balter]{berger2017adaptsize}
D.~S. Berger, R.~Sitaraman, and M.~Harchol-Balter.
\newblock Adaptsize: Orchestrating the hot object memory cache in a content
  delivery network.
\newblock In \emph{USENIX NSDI}, pages 483--498, 2017.

\bibitem[Ch{\l}{\k{e}}dowski et~al.(2021)Ch{\l}{\k{e}}dowski, Polak, Szabucki,
  and {\.Z}o{\l}na]{chlkedowski2021robust}
J.~Ch{\l}{\k{e}}dowski, A.~Polak, B.~Szabucki, and K.~T. {\.Z}o{\l}na.
\newblock Robust learning-augmented caching: An experimental study.
\newblock In \emph{International Conference on Machine Learning}, pages
  1920--1930. PMLR, 2021.

\bibitem[{CISCO}(2019)]{vni2019}
{CISCO}.
\newblock Cisco visual networking index: Forecast and trends 2022, February
  2019.
\newblock Available at
  \url{https://www.cisco.com/c/en/us/solutions/collateral/service-provider/visual-networking-index-vni/white-paper-c11-741490.pdf},
  accessed 09/18/19.

\bibitem[Costa and Pazos(2017)]{costa2017mlcache}
R.~Costa and J.~Pazos.
\newblock Mlcache: A multi-armed bandit policy for an operating system page
  cache.
\newblock Technical report, Technical Report. University of British Columbia,
  2017.

\bibitem[Duplyakin et~al.(2019)Duplyakin, Ricci, Maricq, Wong, Duerig, Eide,
  Stoller, Hibler, Johnson, Webb, Akella, Wang, Ricart, Landweber, Elliott,
  Zink, Cecchet, Kar, and Mishra]{Duplyakin+:ATC19}
D.~Duplyakin, R.~Ricci, A.~Maricq, G.~Wong, J.~Duerig, E.~Eide, L.~Stoller,
  M.~Hibler, D.~Johnson, K.~Webb, A.~Akella, K.~Wang, G.~Ricart, L.~Landweber,
  C.~Elliott, M.~Zink, E.~Cecchet, S.~Kar, and P.~Mishra.
\newblock The design and operation of {CloudLab}.
\newblock In \emph{Proceedings of the {USENIX} Annual Technical Conference
  (ATC)}, pages 1--14, July 2019.
\newblock URL \url{https://www.flux.utah.edu/paper/duplyakin-atc19}.

\bibitem[Einziger and Friedman(2014)]{einziger2014tinylfu}
G.~Einziger and R.~Friedman.
\newblock Tinylfu: A highly efficient cache admission policy.
\newblock In \emph{IEEE Euromicro PDP}, pages 146--153, 2014.

\bibitem[Einziger et~al.(2018)Einziger, Eytan, Friedman, and
  Manes]{einziger2018adaptive}
G.~Einziger, O.~Eytan, R.~Friedman, and B.~Manes.
\newblock Adaptive software cache management.
\newblock In \emph{ACM Middleware}, pages 94--106, 2018.

\bibitem[Eisenman et~al.(2019)Eisenman, Cidon, Pergament, Haimovich, Stutsman,
  Alizadeh, and Katti]{eisenman2019flashield}
A.~Eisenman, A.~Cidon, E.~Pergament, O.~Haimovich, R.~Stutsman, M.~Alizadeh,
  and S.~Katti.
\newblock Flashield: a hybrid key-value cache that controls flash write
  amplification.
\newblock In \emph{USENIX NSDI}, pages 65--78, 2019.

\bibitem[Eytan et~al.(2020)Eytan, Harnik, Ofer, Friedman, and Kat]{eytan2020s}
O.~Eytan, D.~Harnik, E.~Ofer, R.~Friedman, and R.~Kat.
\newblock It's time to revisit $\{$LRU$\}$ vs.$\{$FIFO$\}$.
\newblock In \emph{12th USENIX Workshop on Hot Topics in Storage and File
  Systems (HotStorage 20)}, 2020.

\bibitem[Fan et~al.(2013)Fan, Andersen, and Kaminsky]{fan2013memc3}
B.~Fan, D.~G. Andersen, and M.~Kaminsky.
\newblock {MemC3}: Compact and concurrent memcache with dumber caching and
  smarter hashing.
\newblock In \emph{USENIX NSDI}, pages 371--384, 2013.

\bibitem[Jiang and Zhang(2002)]{jiang2002lirs}
S.~Jiang and X.~Zhang.
\newblock {LIRS}: an efficient low inter-reference recency set replacement
  policy to improve buffer cache performance.
\newblock \emph{ACM SIGMETRICS}, 30\penalty0 (1):\penalty0 31--42, 2002.

\bibitem[Johnson and Shasha(1994)]{johnson19942q}
T.~Johnson and D.~Shasha.
\newblock {2Q}: A low overhead high performance buffer management replacement
  algorithm.
\newblock In \emph{VLDB}, pages 439--450, 1994.

\bibitem[Karedla et~al.(1994)Karedla, Love, and Wherry]{karedla1994caching}
R.~Karedla, J.~S. Love, and B.~G. Wherry.
\newblock Caching strategies to improve disk system performance.
\newblock \emph{IEEE Computer}, 27\penalty0 (3):\penalty0 38--46, 1994.

\bibitem[Ke et~al.(2017)Ke, Meng, Finley, Wang, Chen, Ma, Ye, and
  Liu]{ke2017lightgbm}
G.~Ke, Q.~Meng, T.~Finley, T.~Wang, W.~Chen, W.~Ma, Q.~Ye, and T.-Y. Liu.
\newblock Lightgbm: A highly efficient gradient boosting decision tree.
\newblock In \emph{Advances in Neural Information Processing Systems}, pages
  3146--3154, 2017.

\bibitem[Kirilin et~al.(2020)Kirilin, Sundarrajan, Gorinsky, and
  Sitaraman]{kirilin2020rl}
V.~Kirilin, A.~Sundarrajan, S.~Gorinsky, and R.~K. Sitaraman.
\newblock Rl-cache: Learning-based cache admission for content delivery.
\newblock \emph{IEEE Journal on Selected Areas in Communications}, 38\penalty0
  (10):\penalty0 2372--2385, 2020.

\bibitem[Liu et~al.(2020)Liu, Hashemi, Swersky, Ranganathan, and
  Ahn]{liu2020imitation}
E.~Liu, M.~Hashemi, K.~Swersky, P.~Ranganathan, and J.~Ahn.
\newblock An imitation learning approach for cache replacement.
\newblock In \emph{International Conference on Machine Learning}, pages
  6237--6247. PMLR, 2020.

\bibitem[Lykouris and Vassilvtiskii(2018)]{lykouris2018competitive}
T.~Lykouris and S.~Vassilvtiskii.
\newblock Competitive caching with machine learned advice.
\newblock In \emph{International Conference on Machine Learning}, pages
  3296--3305. PMLR, 2018.

\bibitem[Mattson et~al.(1970)Mattson, Gecsei, Slutz, and Traiger]{mattson}
R.~L. Mattson, J.~Gecsei, D.~R. Slutz, and I.~L. Traiger.
\newblock Evaluation techniques for storage hierarchies.
\newblock In \emph{IBM Systems journal}, volume~9, pages 78--117, 1970.

\bibitem[McAllister et~al.(2021)McAllister, Berg, Tutuncu-Macias, Yang,
  Gunasekar, Lu, Berger, Beckmann, and Ganger]{mcallister2021kangaroo}
S.~McAllister, B.~Berg, J.~Tutuncu-Macias, J.~Yang, S.~Gunasekar, J.~Lu, D.~S.
  Berger, N.~Beckmann, and G.~R. Ganger.
\newblock Kangaroo: Caching billions of tiny objects on flash.
\newblock In \emph{Proceedings of the ACM SIGOPS 28th Symposium on Operating
  Systems Principles}, pages 243--262, 2021.

\bibitem[Megiddo and Modha(2003)]{megiddo2003arc}
N.~Megiddo and D.~S. Modha.
\newblock {ARC}: A self-tuning, low overhead replacement cache.
\newblock In \emph{USENIX FAST}, volume~3, pages 115--130, 2003.

\bibitem[Rohatgi(2020)]{rohatgi2020near}
D.~Rohatgi.
\newblock Near-optimal bounds for online caching with machine learned advice.
\newblock In \emph{Proceedings of the Fourteenth Annual ACM-SIAM Symposium on
  Discrete Algorithms}, pages 1834--1845. SIAM, 2020.

\bibitem[Song et~al.(2020)Song, Berger, Li, Shaikh, Lloyd, Ghorbani, Kim,
  Akella, Krishnamurthy, Witchel, et~al.]{song2020learning}
Z.~Song, D.~S. Berger, K.~Li, A.~Shaikh, W.~Lloyd, S.~Ghorbani, C.~Kim,
  A.~Akella, A.~Krishnamurthy, E.~Witchel, et~al.
\newblock Learning relaxed belady for content distribution network caching.
\newblock In \emph{17th $\{$USENIX$\}$ Symposium on Networked Systems Design
  and Implementation ($\{$NSDI$\}$ 20)}, pages 529--544, 2020.

\bibitem[Tang et~al.(2015)Tang, Huang, Lloyd, Kumar, and Li]{tang2015ripq}
L.~Tang, Q.~Huang, W.~Lloyd, S.~Kumar, and K.~Li.
\newblock {RIPQ}: advanced photo caching on flash for facebook.
\newblock In \emph{USENIX FAST}, pages 373--386, 2015.

\bibitem[Vietri et~al.(2018)Vietri, Rodriguez, Martinez, Lyons, Liu,
  Rangaswami, Zhao, and Narasimhan]{vietri2018lecar}
G.~Vietri, L.~V. Rodriguez, W.~A. Martinez, S.~Lyons, J.~Liu, R.~Rangaswami,
  M.~Zhao, and G.~Narasimhan.
\newblock Driving cache replacement with {ML}-based {LeCaR}.
\newblock In \emph{USENIX HotStorage}, 2018.

\bibitem[Wang et~al.(2019)Wang, He, Alizadeh, and Mao]{wang2019learning}
H.~Wang, H.~He, M.~Alizadeh, and H.~Mao.
\newblock Learning caching policies with subsampling.
\newblock In \emph{NeurIPS Machine Learning for Systems Workshop}, 2019.

\bibitem[Wei(2020)]{wei2020better}
A.~Wei.
\newblock Better and simpler learning-augmented online caching.
\newblock \emph{arXiv preprint arXiv:2005.13716}, 2020.

\bibitem[Yan and Li(2020)]{yan2020rl}
G.~Yan and J.~Li.
\newblock Rl-b{\'e}l{\'a}dy: A unified learning framework for content caching.
\newblock In \emph{Proceedings of the 28th ACM International Conference on
  Multimedia}, pages 1009--1017, 2020.

\bibitem[Yan et~al.(2021)Yan, Li, and Towsley]{yan2021learning}
G.~Yan, J.~Li, and D.~Towsley.
\newblock Learning from optimal caching for content delivery.
\newblock In \emph{Proceedings of the 17th International Conference on emerging
  Networking EXperiments and Technologies}, pages 344--358, 2021.

\bibitem[Yang et~al.(2020)Yang, Yue, and Rashmi]{yang2020large}
J.~Yang, Y.~Yue, and K.~Rashmi.
\newblock A large scale analysis of hundreds of in-memory cache clusters at
  twitter.
\newblock In \emph{14th $\{$USENIX$\}$ Symposium on Operating Systems Design
  and Implementation ($\{$OSDI$\}$ 20)}, pages 191--208, 2020.

\bibitem[Yang et~al.(2021)Yang, Yue, and Vinayak]{yang2021segcache}
J.~Yang, Y.~Yue, and R.~Vinayak.
\newblock Segcache: a memory-efficient and scalable in-memory key-value cache
  for small objects.
\newblock In \emph{18th USENIX Symposium on Networked Systems Design and
  Implementation (NSDI 21)}, pages 503--518, 2021.

\bibitem[Zhou and Maas(2021)]{zhou2021learning}
G.~Zhou and M.~Maas.
\newblock Learning on distributed traces for data center storage systems.
\newblock \emph{Proceedings of Machine Learning and Systems}, 3, 2021.

\end{thebibliography}
